\documentclass[%
reprint,
superscriptaddress,
showpacs,preprintnumbers,
amsmath,amssymb,
aps,
prl,
]{revtex4-2}

\usepackage{color}
\usepackage{graphicx}
\usepackage{bm}
\usepackage{mathrsfs}
\usepackage{subfigure}
\usepackage{float}
\usepackage{dsfont}
\usepackage{ulem}
\usepackage[colorlinks,allcolors=blue]{hyperref}

\def\bra#1{\langle #1|}
\def\ket#1{ |#1 \rangle}

\begin{document}

\title{Experimental demonstration of Quantum Overlapping Tomography}
\author{Yang Zhengning}
\affiliation{Division of Physics and Applied Physics, School of Physical and Mathematical Sciences, Nanyang Technological University, Singapore 637371, Singapore}
\author{Shihao Ru}
\affiliation{Division of Physics and Applied Physics, School of Physical and Mathematical Sciences, Nanyang Technological University, Singapore 637371, Singapore}
\affiliation{School of Physics, Xi'an Jiaotong University, Xi'an 710049, China}
\author{Lianzhen Cao}
\affiliation{School of Physics and Photoelectric Engineering, Weifang University, Weifang 261061, China}
\author{Nikolay Zheludev}
\affiliation{Division of Physics and Applied Physics, School of Physical and Mathematical Sciences, Nanyang Technological University, Singapore 637371, Singapore}
\affiliation{The Photonics Institute and Centre for Disruptive Photonic Technologies, Nanyang Technological University, Singapore 637371, Singapore}
\author{Weibo Gao}
\email{wbgao@ntu.edu.sg}
\affiliation{Division of Physics and Applied Physics, School of Physical and Mathematical Sciences, Nanyang Technological University, Singapore 637371, Singapore}
\affiliation{The Photonics Institute and Centre for Disruptive Photonic Technologies, Nanyang Technological University, Singapore 637371, Singapore}
\affiliation{Center for Quantum Technologies, National University of Singapore, Singapore 117543, Singapore}

\date{\today}

\begin{abstract}
Quantum tomography is one of the major challenges of large-scale quantum information research due to the exponential time complexity. In this work, we develop and apply a Bayesian state estimation method to experimentally demonstrate quantum overlapping tomography [Phys. Rev. Lett. \textbf{124}, 100401 (2020)], a scheme intent on characterizing critical information of a many-body quantum system in logarithmic time complexity. By comparing the measurement results of full state tomography and overlapping tomography, we show that overlapping tomography gives accurate information of the system with much fewer state measurements than {full state tomography}.
\end{abstract}


\maketitle


As more interest and effort have been put into the research of quantum information processing in recent years \cite{Ladd2010,nielsen_chuang_2010}, there have been remarkable advances in constructing and controlling large-scale quantum systems with a series of physical systems, including but not limited to superconducting circuits \cite{Schoelkopf2013, Arute2019, PhysRevLett.127.180501}, linear optics \cite{PhysRevLett.127.180502, RevModPhys.79.135}, ion trap \cite{Monroe2013, Figgatt2019}, and ultracold atoms \cite{RevModPhys.82.2313}. Although it has been realistic to create and operate an extensive system with around 100 or even 1000 qubits \cite{Yang2020, Yang2020Science}, it’s still a question of how to measure such many-body states and demonstrate the correlation between any two parts of the system. Due to the quantum nature of qubits, the information carried by a qubit cannot be read out with one single measurement \cite{1983LNP}. Instead, one needs to perform multiple times of measurement with multiple sets of basis on one quantum state to reconstruct the density matrix representing the state \cite{PhysRevLett.86.4195}. As the number of qubits in the system goes up, the number of required measurements increases exponentially \cite{ODonnell2016}, leading to an unacceptable time complexity, which could overwhelm the system's stability for even a moderate scale. In fact, for a system with just 10 qubits, a full state tomography (FST) has been considerably hard \cite{PhysRevLett.119.180511}. Driven by this challenge, various protocols have been raised to reduce the time complexity. Some protocols offer advantages for certain quantum states with special structures \cite{Lanyon2017}. Some protocols can estimate an unknown state with higher efficiency, but they require quantum non-demolition measurement, which remains experimentally unavailable nowadays \cite{Aaronson2020}. 

\begin{figure}[hb]
\centering
\includegraphics[width=\linewidth]{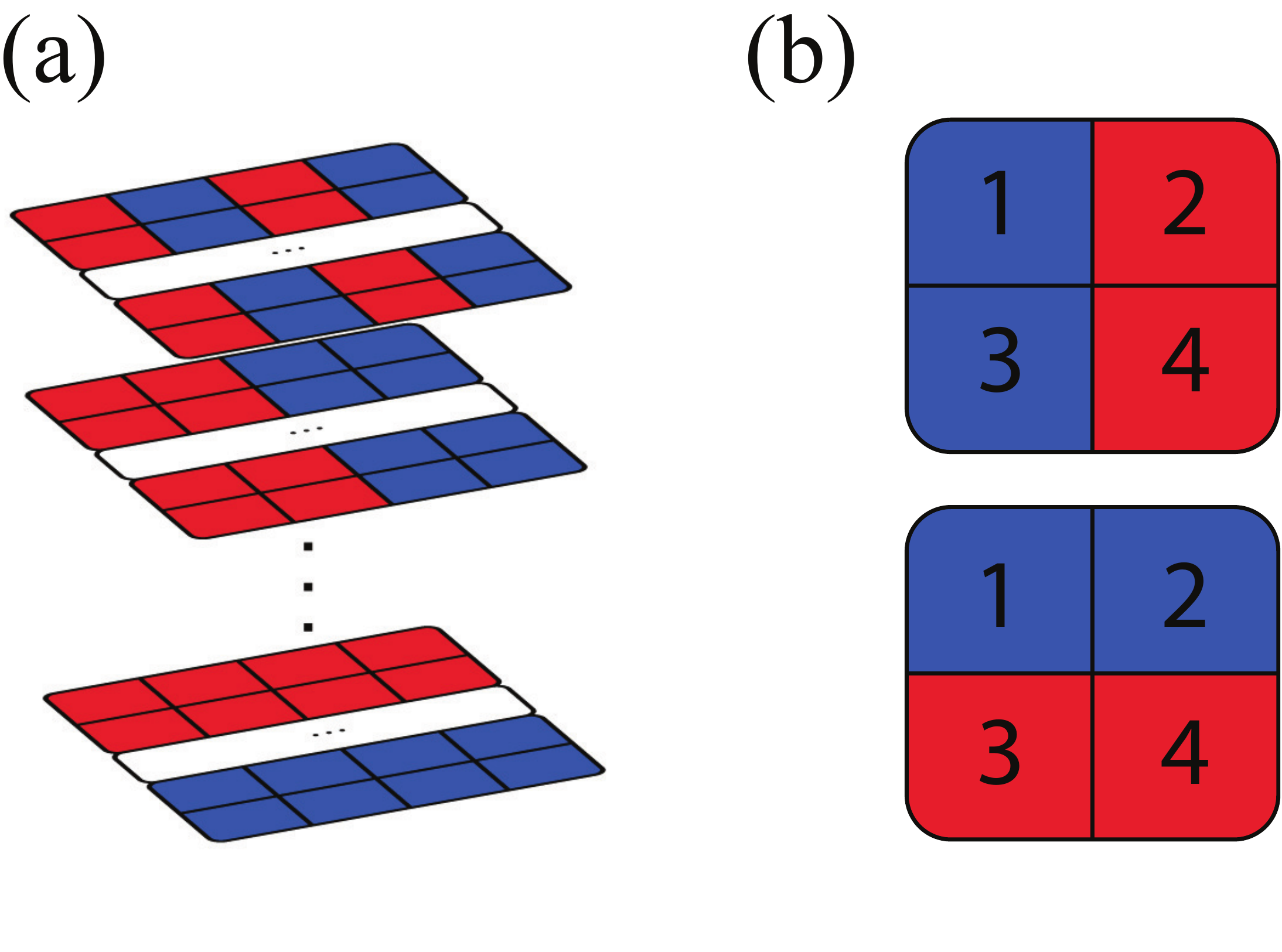}
\caption{(a) 2-qubit quantum overlapping tomography of a large-scale system. The whole system is divided into two groups, red and blue, in different strategies. For each dividing strategy, the two groups are measured on a different basis. (b) QOT dividing strategy for the $n=4, k=2$ case.}\label{fig:1}
\end{figure}

A more realistic idea is to retrieve limited but critical information by reconstructing the reduced density matrices of the small-scale subsystems of the huge-scale system, with much fewer measurements. Although this kind of ‘partial’ tomography does not give a complete picture of the system, it is usually critical in application cases, such as the research on long-range order in many-body systems \cite{RevModPhys.77.259} and machine learning based on quantum neural networks \cite{Schuld2014}. However, even for this simplified task, the time complexity still can be unacceptable. Supposing we have a system of n qubits and hope to measure all the k-qubit subsystems, there are $\binom{n}{k}$ subsystems that need to be measured. For small k relative to n, we have $\binom{n}{k} \sim n^k$. Thus, the time complexity is $e^{O(k)}\times n^k$ because measuring each k-qubit subsystem has a time complexity $e^{O(k)}$. Even for a modest scale of $n=50$ and $k=2$, the total measurements will reach ~$10000N$, where $N$ is the required number of measurements to obtain a statistically significant result for each measurement setting. This number has been too much for an actual experiment.

Quantum Overlapping Tomography (QOT) \cite{PhysRevLett.124.100401} is one protocol proposed by J. Cotler and F. Wilczek, which makes use of the strong power of parallelism to simplify this task. With the overlapping nature of quantum many-body systems, QOT can reconstruct all $\binom{n}{k}$ k-qubit subsystems in logarithmic time complexity $\sim e^{O(k)}\log_{k}{n} $ by reorganizing the measurement dataset. With this scheme, for $n=50,$ $k=2$, it’s possible to reconstruct all the subsystems within less than 100N measurements, which means an over 100 times boost.

{Due to its fascinating advantage, QOT has been attracting considerable attention from the community. Some researchers have started to use this method in their projects \cite{Maciejewski2021modelingmitigation}. However, a general experimental scheme of QOT and the performance comparison between QOT and FST are still new topics to fill in.} In this work, we experimentally demonstrate QOT on a 4-photon polarization entanglement system powered by spontaneous parametric down-conversion (SPDC).
To apply the QOT scheme originating from exact tomography \cite{ALTEPETER2005105}, which is not legitimate to use in experiments, into the experiment scenario, we develop an algorithm to perform a Bayesian mean estimation (BME) \cite{Blume_Kohout_2010} to estimate the target states with the measurement dataset, for both FST and QOT.
We compare the outcome of FST and QOT with a 4-photon GHZ state. To make the results comparable, the QOT estimation is performed with a subset of the measurement dataset that the FST uses instead of a separate dataset. In addition, we generate a variation of GHZ state and show that QOT can characterize general kinds of states.

In quantum information, any quantum state of a single qubit can be demonstrated with a two-dimension density matrix \cite{ALTEPETER2005105}
\begin{align}
\hat{\rho}=\frac{1}{2} \sum_{i=0}^{3} S_{i} \hat{\sigma}_{i}, 
\hat{\sigma}_0\equiv 
\begin{pmatrix}
 1 & 0 \\
 0 & 1 \\
\end{pmatrix}, 
\hat{\sigma}_1\equiv 
\begin{pmatrix}
 0 & 1\\
 1 & 0
\end{pmatrix}, \notag\\
\hat{\sigma}_2\equiv 
\begin{pmatrix}
 0 & -i\\
 i & 0
\end{pmatrix}, 
\hat{\sigma}_3\equiv 
\begin{pmatrix}
 1 & 0\\
 0 & -1
\end{pmatrix}
\end{align}

$S_i$ values can be given by $S_i=Tr\left \{ \hat{\sigma_i}\hat{\rho } \right \} $, which indicates that it can be directly obtained with projective measurements. 

Similarly, a general multi-qubit system with n qubits can be demonstrated with a density matrix with $2^n$ dimensions.

\begin{equation}
\hat{\rho}=\frac{1}{2^n}\sum_{i_1 ,i_2, \cdots,i_n =0}^{3} S_{i_1 ,i_2, \cdots,i_n} \hat{\sigma}_{i_1}\otimes\hat{\sigma}_{i_2}\otimes\cdots\hat{\sigma}_{i_n} 
\end{equation}

To reconstruct the density matrix, the main task of a {quantum full state tomography} is to obtain all the $S_{i_1 ,i_2, \cdots , i_n}$ values.
{In theory, the values can be directly calculated with the results of measurement through exact tomography \cite{ALTEPETER2005105}}.

\begin{widetext}
\begin{align}
S_{i_1 ,i_2, \cdots,i_n} &= (\lambda_1P_{\phi_{i_1}}+\lambda_1^\bot P_{\phi_{i_1}^\bot })(\lambda_2P_{\phi_{i_2}}+\lambda_2^\bot P_{\phi_{i_2}^\bot })\cdots(\lambda_nP_{\phi_{i_n}}+\lambda_n^\bot P_{\phi_{i_n}^\bot }) \notag \\
&=(\lambda_1\lambda_2\cdots\lambda_n)P_{\phi_{i_1}\phi_{i_2}\cdots\phi_{i_n}}+(\lambda_1\lambda_2\cdots\lambda_n^\bot)P_{\phi_{i_1}\phi_{i_2}\cdots\phi_{i_n^\bot}}+\cdots+(\lambda_1^\bot\lambda_2^\bot\cdots\lambda_n^\bot)P_{\phi_{i_1^\bot}\phi_{i_2^\bot}\cdots\phi_{i_n^\bot}}
\end{align}
\end{widetext}
Here we note $\phi_{i_j}$ as the eigenstate with eigenvalue $\lambda_j=1$ and $\phi_{i_j}^\bot$ with eigenvalue $\lambda_j^\bot=-1. P_{\phi_{i_j}^\bot}$ stands for the {probability}. that $j^{th}$ qubit is in $\phi_{i_j}^\bot$, which can be estimated with a finite number of measurements. For the case of $i_j=0$, $\lambda_jP_{\phi_0} + \lambda_j^\bot P_{\phi_0^\bot } =1$, which means this term would be ``transparent" in the calculation.

For the whole system with $n$ qubits, an array with $2n$ detectors is deployed to measure all $2^n$ eigenstates of element density matrix $ \hat{\sigma}_{i_1}\otimes\hat{\sigma}_{i_2}\otimes\cdots\hat{\sigma}_{i_n}$ at the same time, with the measurement setting $\left \{i_1 ,i_2 , \cdots , i_n \right \}$. In principle, $4^n$ settings are needed since each $i_j$ has 4 possible values. However, for those settings with any $i_j=0$, the $S_{i_1 ,i_2, \cdots,i_n}$ can be directly calculated with the measurement results by other settings with all $i_j=0$. Thus, in practice, we need $3^n$ settings to reconstruct the density matrix of a system with $n$ qubits.

Note that exact tomography is only usable with the assumption that the observed probabilities are theoretically perfect, which means the observation should be with no errors and be conducted through an infinite ensemble of states \cite{ALTEPETER2005105}. In other word, Eq(3) is not valid for any actual measurement, otherwise, it would lead to physically insufficient results. A common practice to reconstruct legitimate density matrices with real observation results is statistical estimation methods, such as Maximum Likelihood Estimation (MLE) \cite{PhysRevA.61.010304} and Bayesian Mean Estimation (BME) \cite{Blume_Kohout_2010, Lukens_2020}. In this work, we develop an algorithm based on Gibbs sampling \cite{4767596} to perform a Bayesian Mean Estimation with the measurement datasets for both FST and QOT.

Here we show how the QOT works in a $k=2$ case. Firstly, we describe the task as reconstructing all $\binom{n}{k}$ 2-qubit reduced density matrices:
\begin{equation}
\hat{\rho}^{\left \{ x_1,x_2 \right \} } = \frac{1}{2^2} \sum_{i_1,i_2=0}^{3} S_{i_1,i_2}^{\left \{ x_1,x_2 \right \} }\hat{\sigma}_{i_1}\otimes \hat{\sigma}_{i_2}
\end{equation}
where $\left \{ x_1,x_2 \right \} $ represents a 2-qubit subsystem of the n-qubit system. To obtain the values of $S_{i_1,i_2}^{\left \{ x_1,x_2 \right \} }$, normally we need to pick all of the 2-qubit groups $\left \{ x_1,x_2 \right \} $ and measure them individually. For the $n$-qubit system, there are $\frac{n(n-1)}{2} $ of such subsystems, and $N\times 3^2=9N$ measurements need to be operated on each subsystem to get a tomography of them, which gives an $O(n^2)$ time complexity.

In QOT (Fig.\ref{fig:1}(a)) instead, we divide the n-qubit system into 2 groups, by $q=\left \lceil \log_{2}{n} \right \rceil $ ways. The dividing strategy should satisfy the requirement that for any subsystem $\left \{ x_1,x_2 \right \} $, and there is at least one divide where $x_1$ and $x_2$ are in different groups. 

A dividing example of $n=4$ case is given in (Fig.\ref{fig:1}(b)). The 4-qubit system $\left \{1,2,3,4\right \}$, is divided in 2 different ways noted as $\left \{\left \{1,2\right \},\left \{3,4\right \}\right \}$ and $\left \{\left \{1,3\right \},\left \{2,4\right \}\right \}$. Then we measure the system in two steps:

\begin{itemize}
\item[1.] Measure all the qubits in $X,Y,Z$ basis respectively, which need $3N$ measurements in total.
\item[2.] For each divide out of the $q$ divides, measure all qubits in group 1 in one basis $B_1 \in \left \{ X,Y,Z \right \} $ and measure all qubits in group 2 in another different basis $B_2$. Thus, it takes 6 measurements for each of the q divides, which is a total of $6qN$.
\end{itemize}

So, we use $3+6q$ measurement basis sets in total, which gives a logarithmic time complexity. {With an ideal probabilities dataset,} density matrices $\hat{\rho}^{\left \{ x_1,x_2 \right \} }$ can be reconstructed by calculating the value
\begin{equation}
S_{i_1 ,i_2}^{\left \{ x_1,x_2 \right \} }=(\lambda_1P_{\phi_{i_1}}+\lambda_1^\bot P_{\phi_{i_1}^\bot })(\lambda_2P_{\phi_{i_2}}+\lambda_2^\bot P_{\phi_{i_2}^\bot }).
\end{equation}
{This calculation relies on the same assumption as exact tomography, so it is not valid for real measurements. Proper estimation is also necessary to reconstruct legitimate density matrices with the QOT dataset \cite{supp1}.}

In our experiment, we build up a 4-qubit entangled system and try to reconstruct the 2-qubit subsystems of it \cite{supp1}. If we {operate FST on the 4-qubit system}, we need $N\times 3^4=81N$ measurements for the task. For the subsystem reconstruction task that we have discussed, $N\times\binom{4}{2}\times 3^2=54N$ measurements are required. However, with QOT, only $N\times(3+6\log_{2}{4})=15N$ measurements are required to reconstruct all the 2-qubit subsystems. 

\begin{figure}[h]
\centering
\includegraphics[width=0.92\linewidth]{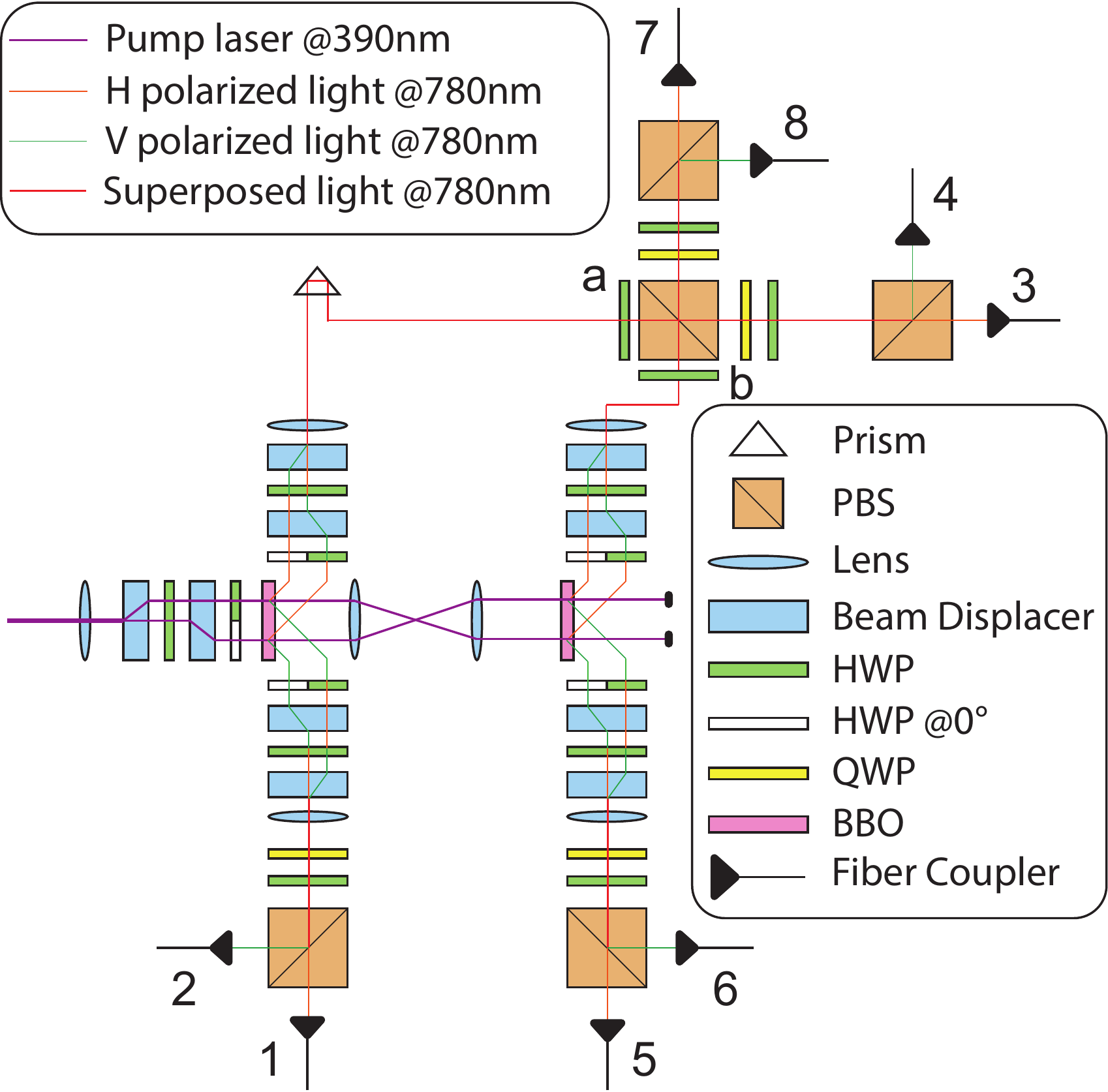}
\caption{Schematic of experimental set-up for generating 4-photon entanglement, with the detectors labeled by order. An ultrafast pulsed laser with the center wavelength of 390 nm, pulse duration of $\sim$100 fs, and repetition rate of 80 MHz was deployed to pump two sets of interference-based beam-like SPDC entanglement sources \cite{PhysRevLett.121.250505,PhysRevLett.75.4337}. The polarization-entangled photon pairs go through a post-selection interference to generate a 4-photon GHZ state, which is then measured by 8 single-photon detectors. To reduce the loss of two-fold fidelity caused by time-space correlation \cite{PhysRevA.64.063815,PhysRevLett.117.210502}, we applied narrow-band filters with $\lambda_{FWHM}$=3 nm and $\lambda_{FWHM}$=10 nm to the signal and idler photons respectively. The center wavelengths of both signal and idler photons are 780 nm. BBO: Barium Borate; PBS: polarization beam-splitter; HWP: half-wave plate; QWP: quarter-wave plate.}\label{fig:2.1}
\end{figure}

\begin{figure}[ht]
\centering
\includegraphics[width=0.95\linewidth]{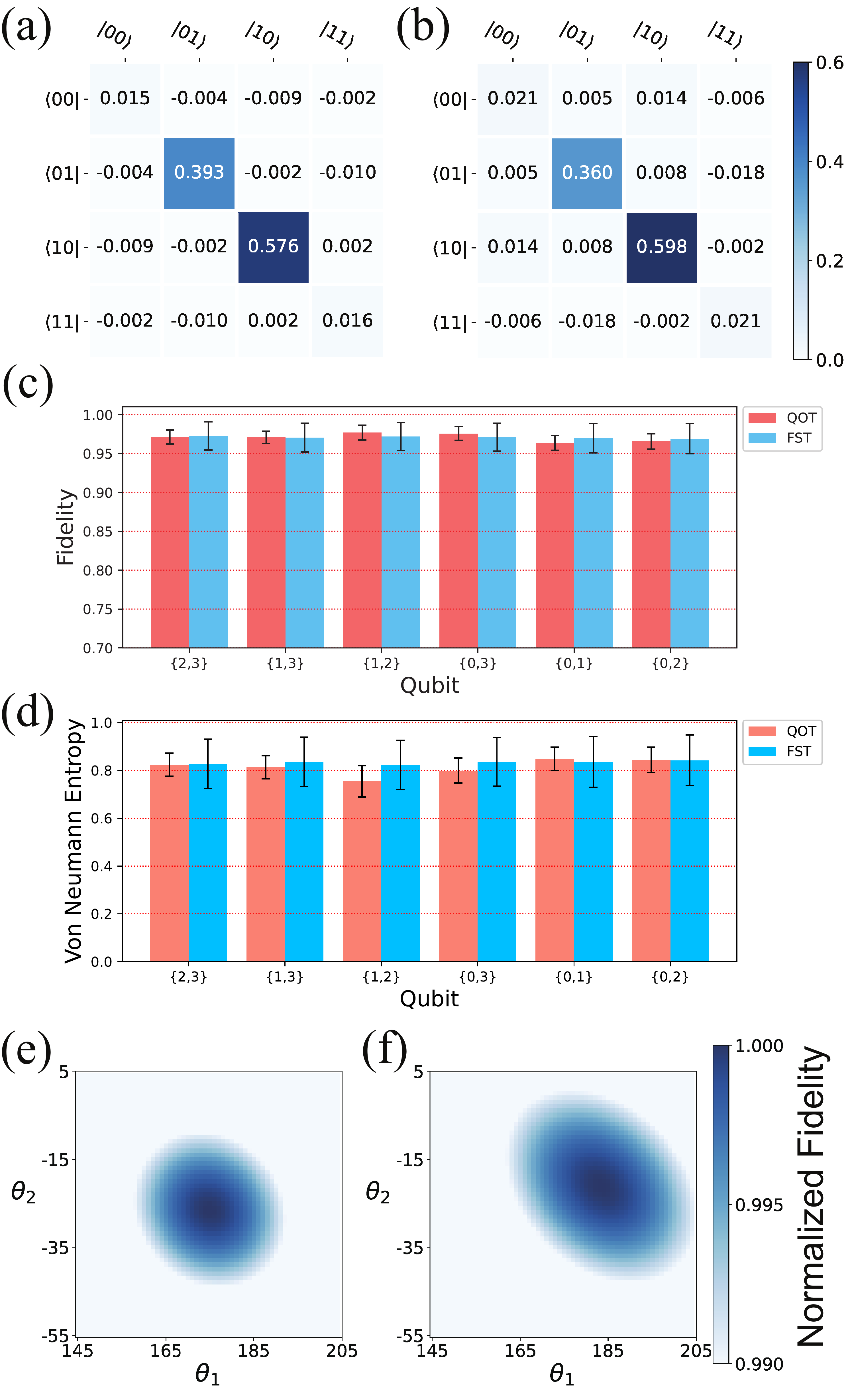}
\caption{Real part of density matrix of 2-qubit subsystem (a) ${\psi}_{2,3}^F$ obtained by 4-qubit full state tomography (FST) (b) ${\psi}_{2,3}^O$ obtained by overlapping tomography (QOT). (c) 2-qubit state fidelities of reconstructed ${\psi}_{i_1,i_2}$ and reference state $(\psi_{GHZ}^R)_{i_1,i_2}$ and (d) Von Neumann Entropy of reconstructed ${\psi}_{i_1,i_2}$. (e) Normalized fidelity $F(\theta_1,\theta_2)$ between 4-qubit state ${\psi'}^F$ reconstructed via {4-qubit FST} and reference state ${\psi'}^R(\theta_1,\theta_2)$, $F$ reached peak at $\theta_1=175^\circ, \theta_2=-27^\circ$. (f) Normalized fidelity $F_{0,2}(\theta_1,\theta_2)$ between 2-qubit subsystem ${\psi'}_{0,2}$ reconstructed via overlapping tomography and reduced reference state ${\psi'}_{0,2}^R(\theta_1,\theta_2)$. $F_{0,2}$ reached peak at $\theta_1=183^\circ, \theta_2=-21^\circ$. For (c-d), QOT and FST results are estimated by Bayesian Estimation based on equal number of measurements. Error bars show 95\% confidence interval obtained by Bayesian analysis based on Markov Chain Monte Carlo method, as 95\% MCMC sample points (after pre-heating) lie within the intervals.}\label{fig:3}
\end{figure}

Fig.\ref{fig:2.1} shows an overview of the experimental set-up for generating and detecting a 4-photon GHZ state $\psi_{GHZ}^R = \frac{1}{\sqrt{2}} (\ket{HVHV} + e^{i\theta}\ket{VHVH})$. Following Jones Calculus \cite{Jones:41}, the horizontal polarization $\ket{H}$ and vertical polarization $\ket{V}$ are defined as the eigenstates of Pauli matrix $\sigma_z$. $\ket{L/R}= \frac{1}{\sqrt{2}}(\ket H\pm i\ket V), \ket{D/A}= \frac{1}{\sqrt{2}}(\ket H\pm \ket V)$ are the eigenstates for $\sigma_y$ and $\sigma_x$ respectively. We use waveplates and polarization beamsplitters to measure qubits on a specific measurement basis. By detecting photons in both output modes of PBS, we measure all 16 probabilities in parallel for one measurement basis set.

The two-fold coincidence counting rate can reach $100$ kHz at pump power $550$ mW, with a state fidelity $F>0.97$. The singles count rates vary between 350 - 500 kHz with different channels, generating a 2-fold accidental rate around 2 kHz and a coincidence-to-accidental ratio (CAR) around 50:1. The 4-fold accidental rate is around 0.05 Hz. With narrow-band filtering, the single photon detecting efficiency is {between $20-25\%$}. We recorded 4-fold coincidences (for 16 sets of measurement basis spontaneously) for 300 s on each setting, with the total coincidence counting rate over 16 sets of basis around 10 Hz{, with a CAR around 200:1}. We performed {an FST} on this state to reconstruct the 4-qubit density matrix $\rho^F$. Comparing the tomography result with a given pure state $\rho^R=\ket{\psi_{GHZ}^R}\bra{\psi_{GHZ}^R}$, the state fidelity $F^M=(Tr\sqrt{\sqrt{\rho^R}\rho^F\sqrt{\rho^R} })^2 $ ={0.922$\pm$0.013} .

For the QOT case, we focus on the 2-qubit subsystems of the 4-qubit entangled state. Thus, firstly, we obtain the density matrices of 2-qubit subsystems $\rho_{j_1,j_2}^F={Tr}_{i_1,i_2}\rho^M, i_1,i_2\in\left \{ 1,2,3,4 \right \}, \left \{ j_1,j_2 \right \}=\left \{ 1,2,3,4 \right \}/\left \{ i_1,i_2 \right \}$ by calculating the partial trace. Then we perform overlapping tomography on the same state, reconstructing the density matrices $\rho_{i_1,i_2}^O$ of 2-qubit subsystems. By comparing $\rho_{i_1,i_2}^F$ and $\rho_{i_1,i_2}^O$ in density matrix visualization (Fig.\ref{fig:3}(a-b)), we confirmed that {QOT} gives highly similar results to the outcome from 4-qubit FST. In Fig.\ref{fig:3}(c), we compare the state fidelity obtained by FST and QOT. The fidelity differences are less than 0.01 within the margin of error. To further characterize the states in this case, we calculated and compared the Von Neumann Entropy (VNE) \cite{PasqualeCalabrese_2004} $S=-Tr(\rho \cdot \ln \rho)$ of $\rho_{i_1,i_2}$ reconstructed with QOT and FST. For the 2-qubit subsystems of an ideal 4-qubit GHZ state, the quantum state should be a mixed state consisting of two maximally entangled Bell states $\ket\Psi$, and the VNE should be $S\approx ln2=0.69$. A lower state fidelity caused by imperfect state generation and measurement usually leads to a higher VNE as well, which indicates the state is ``more mixed" and ``less entangled". By comparing VNE $S_{i_1,i_2}^O$ and $S_{i_1,i_2}^O$ (Fig.\ref{fig:3}(d)), we find that overlapping tomography gives overall similar results with the {4-qubit FST}. Though all differences between VNE results given by QOT and 4-qubit FST are within the margin of error, for some subsystems, the VNE by QOT is visibly smaller than that by FST due to system inconsistency.

Comparing QOT and FST results, the error bars in Fig.\ref{fig:3}(c-d) show that the 95\% confidence interval of both state fidelity and VNE acquired by FST is significantly larger than by QOT. It indicates that QOT gives a better estimate with equal number of measurements. Therefore, QOT would take less measurement time (as measuring fewer copies of the state) to obtain equal error margin. This feature can be a significant advantage of QOT when measuring large-scale systems as QOT takes much less time and suffer less from system inconsistency.

Further, we use overlapping tomography to characterize an alternative state $\psi'$. The reference state wavefunction for this case is ${\psi'}^R=\frac{1}{\sqrt{2}} (\ket{D'VD''V}+\ket{A'HA''H})$, where $\ket{D'}=\ket H+e^{i\theta_1}\ket V, \ket{A'}=\ket H-e^{i\theta_1}\ket V, \ket{D''}=\ket H+e^{i\theta_2}\ket V, \ket{A''}=\ket H-e^{i\theta_2}\ket V$. We generated this state by {setting HWP a and b in Fig.\ref{fig:2.1} to $\theta_a=-22.5^\circ$ and $\theta_b=-22.5^\circ$}. Similarly, we compare the result of 4-qubit FST and QOT with density matrices visualization \cite{supp1}. This result shows QOT is equally efficient for different multi-qubit states, indicating that QOT is a promising method to characterize common quantum states.

In the wavefunction of ${\psi'}^R$, $\theta_1$ and $\theta_2$ are extra phases introduced by the SPDC process, which can be estimated by calculating the fidelity between the measurement result and reference states. We estimated $\theta_1=175^\circ, \theta_2=-21^\circ$ by comparing the 4-qubit density matrix ${\rho'}^F$ and obtained them by {4-qubit FST} with reference state ${\rho'}^R(\theta_1,\theta_2)=\ket{{\psi'}^R(\theta_1,\theta_2)}\bra{{\psi'}^R(\theta_1,\theta_2)}$ (Fig.\ref{fig:3}(e)). We also use the subsystems reconstructed via QOT to estimate the state phase $\theta_1,\theta_2$ by comparing the density matrices ${\rho'}^O_{i_1,i_2}$ with reference state ${\rho'}^R_{i_1,i_2}(\theta_1\theta_2)$ (Fig.\ref{fig:3}(f)). Though a difference exists between the two estimations, considering the small derivative of fidelity to both $\theta$s, the results are reasonably close to each other as the difference between the two estimations corresponds to less than 0.8\% difference in state fidelity. System inconsistency may also contribute to the difference. This result shows that QOT can extract important state parameters with a significantly reduced time complexity. 

\begin{figure}[h]
\centering
\includegraphics[width=\linewidth]{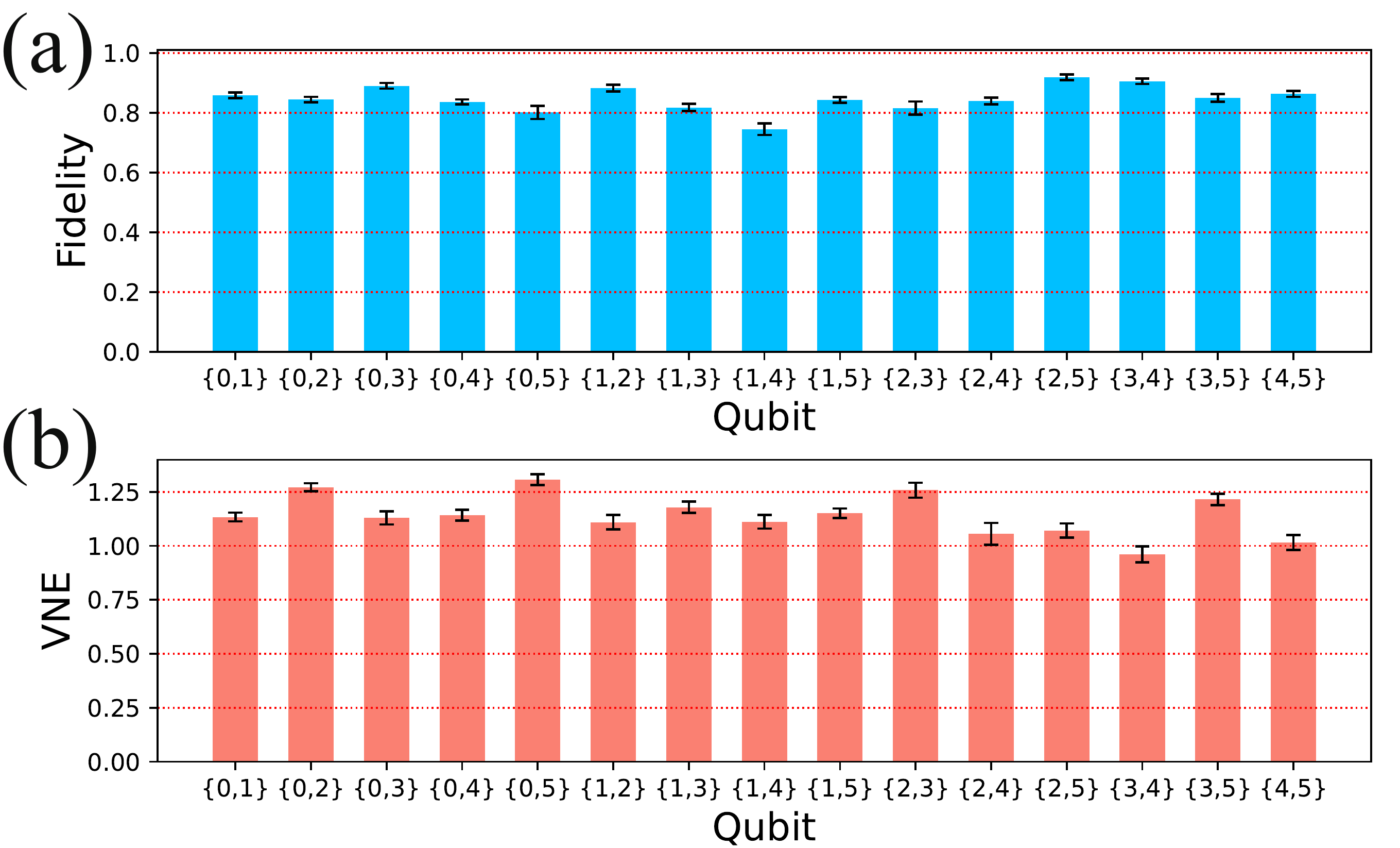}
\caption{2-qubit state (a) fidelity with subsystem of 6-photon GHZ state $({\psi}_{{GHZ}_6}^R)_{i_1,i_2}$ and (b) Von Neumann Entropy of reconstructed density matrix ${\psi}_{i_1,i_2}$ obtained by 6-qubit QOT, error bars show 95\% confidence interval obtained by Bayesian analysis based on MCMC method, as 95\% sample points lie within the intervals.}\label{fig:4}
\end{figure}

To give a further demonstration of the advantage of QOT, we perform QOT on a linear optical six-qubit system \cite{supp1}. We attempt to genearate the state ${\psi}_{{GHZ}_6}^R= \frac{1}{\sqrt{2}} (\ket{HVHVHV} + \ket{VHVHVH})$ and perform QOT on it. Six-photon coincidence events are observed at a count rate of 0.05 Hz. Fig.\ref{fig:4} shows the state fidelity and the VNE obtained in this experiment. To perform a six-qubit FST, the state needs to be measured in 729 sets of basis. As 700 events are recorded for each basis sets to get a reasonable witness, the whole FST would take around 120 days, which is unacceptable given the stability of optical set-ups. By contrast, with QOT, we can reconstruct all 15 2-qubit subsystems with only 21 sets of basis, and the measurement only take around 80 hours. 

In summary, we performed quantum overlapping tomography to characterize 4-qubit GHZ states. By comparing the QOT results with {FST} results, we show that QOT can reconstruct subsystems with a given scale and extract important parameters of a 4-qubit state with a remarkably reduced number of measurements (or give a better estimation with equal number of measurements). This scheme is directly applicable to quantum states with a broader range of structures and larger scale, which could be significant for the future development of quantum information.

We acknowledge Singapore National Research foundation through QEP grant (NRF2021-QEP2-01-P02, NRF2021-QEP2-03-P01, 2019-0643 (QEP-P2) and 2019-1321 (QEP-P3)) and Singapore Ministry of Education (MOE2016-T3-1-006 (S)).




%

\end{document}


\title{Supplementary Information}


\date{\today}


\maketitle



\renewcommand{\thefigure}{S\arabic{figure}}
\renewcommand{\theequation}{S\arabic{equation}}
\renewcommand{\thetable}{S\arabic{table}}

\section{Definition and Notation of the photon polarization qubits}
We follow Jones Calculus \cite{Jones:41} to define the photon polarization qubits. (Table \ref{tab:1}) Following this definition, the eigenstates of Pauli matrix $\sigma_x$ are $\ket{D}$ and $\ket{A}$. Similarly, we have eigenstates $\ket{L}$ and $\ket{R}$ for $\sigma_y$ and $\ket{H}$ and $\ket{V}$ for $\sigma_z$.

\begin{table}[h]
\caption{Notation definition of photon polarization qubit}

\begin{tabular}{ c | c | p{4.7cm} < {\centering} }
\hline
\hline
State Vector &
Ket Notation &
Polarization\\
\hline
$\begin{pmatrix}
 1\\0
\end{pmatrix}$ & $\ket{H}$ & Linear polarization in x-direction\\
\hline
$\begin{pmatrix}
 0\\1
\end{pmatrix}$ & $\ket{V}$ & Linear polarization at $90^{\circ}$ from x-direction\\
\hline
$\displaystyle{\frac{1}{\sqrt{2}}}\begin{pmatrix}
 1\\1
\end{pmatrix}$ & $\ket{D}$ & Linear polarization at $45^{\circ}$ from x-direction\\
\hline
$\displaystyle{\frac{1}{\sqrt{2}}}\begin{pmatrix}
 1\\-1
\end{pmatrix}$ & $\ket{A}$ & Linear polarization at $-45^{\circ}$ from x-direction\\
\hline
$\displaystyle{\frac{1}{\sqrt{2}}}\begin{pmatrix}
 1\\-i
\end{pmatrix}$ & $\ket{R}$ & Right-hand circular polarization\\
\hline
$\displaystyle{\frac{1}{\sqrt{2}}}\begin{pmatrix}
 1\\i
\end{pmatrix}$ & $\ket{L}$ & Left-hand circular polarization \\
\hline 
\hline
\end{tabular}
\label{tab:1}
\end{table}

\begin{figure}[b]
\centering
\includegraphics[width=0.5\linewidth]{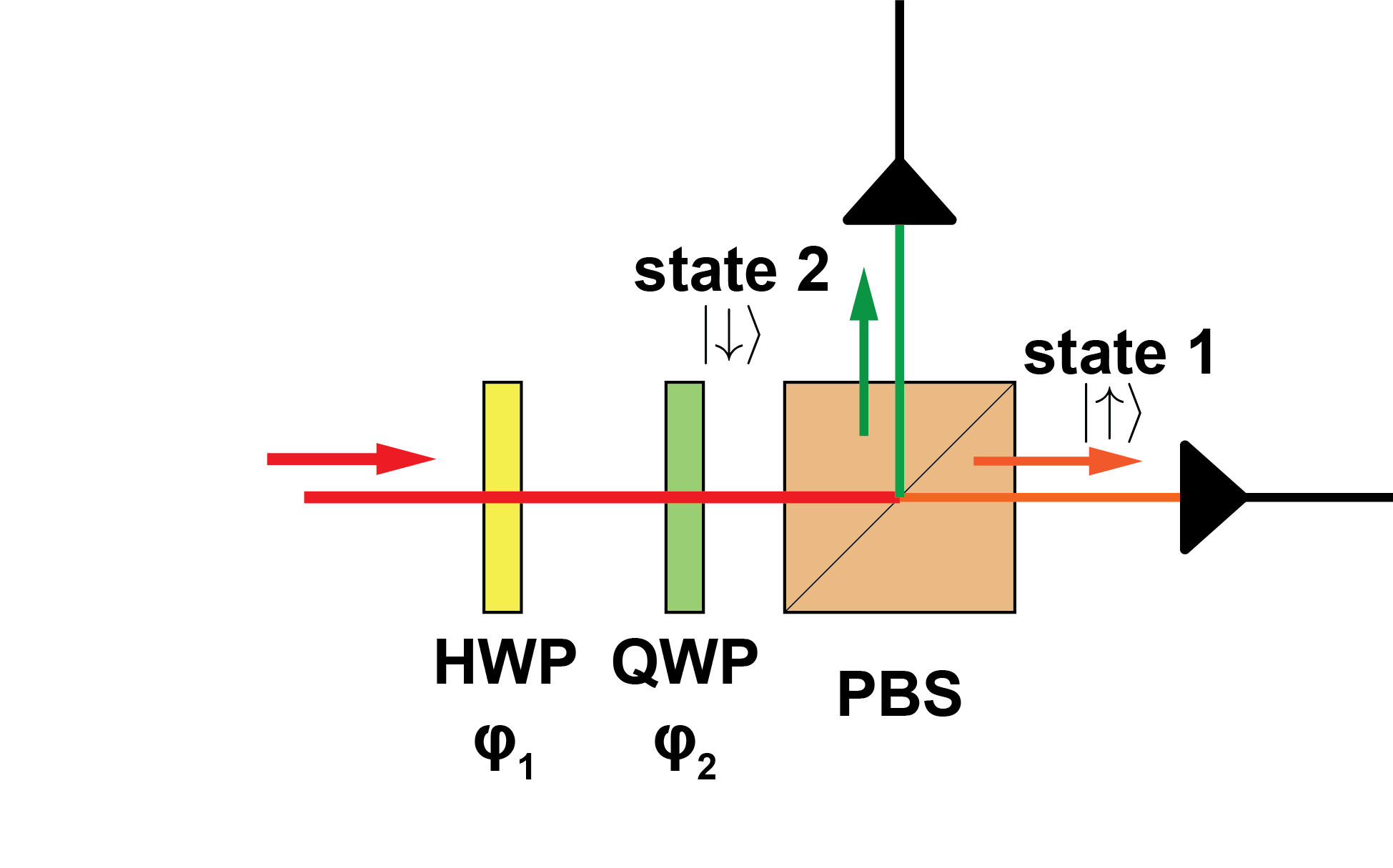}
\caption{Photon detection unit to measure a single polarized photon qubit. The combination of a half-wave plate(HWP), a quarter-wave plate(QWP), and a polarization beam-splitter(PBS) can project a polarized photon to any given basis.}\label{fig:1}
\end{figure}

\section{Experimental State Measurement of photon polarization qubit.}
Fig.\ref{fig:1} shows the optical detection unit to measure a single polarized photon qubit. The waveplates are set to particular (Table \ref{tab:2}) angles to project the state on a certain measurement basis. The PBS seperates two eigenstates spatially and sends photons on different eigenstates to different detectors. By identifying the recorded event on each detector, we count the qubits in the corresponding state.

\begin{table}[h]
\caption{Measurement setting of polarized single photons. The HWP is set at $\phi_1$, the QWP is set at $\phi_2$, state 1(also noting as $\ket{\uparrow}$) corresponds to the transmission mode of PBS, and state 2 ($\ket{\downarrow}$) corresponds to the reflection mode.}
\begin{tabular}{ p{2cm} < {\centering} | p{2cm} < {\centering} | p{1.5cm} < {\centering} | p{1.5cm} < {\centering} | c }
\hline
\hline
$\phi_1$ & $\phi_2$ & State 1 & State 2 & Basis Notation\\
\hline
$0^{\circ}$ & $0^{\circ}$ & $\ket{H}$ & $\ket{V}$ & Z\\
${22.5}^{\circ}$ & $0^{\circ}$ & $\ket{D}$ & $\ket{A}$ & X\\
$0^{\circ}$ & ${45}^{\circ}$ & $\ket{R}$ & $\ket{L}$ & Y \\
\hline
\hline
\end{tabular}
\label{tab:2}
\end{table}

\section{Experimental Methodology of 4-qubit Full Tomography}
\subsection{State Measurement of 4-qubit Photon Polarization System}

\begin{figure}[h]
\centering
\includegraphics[width=0.5\linewidth]{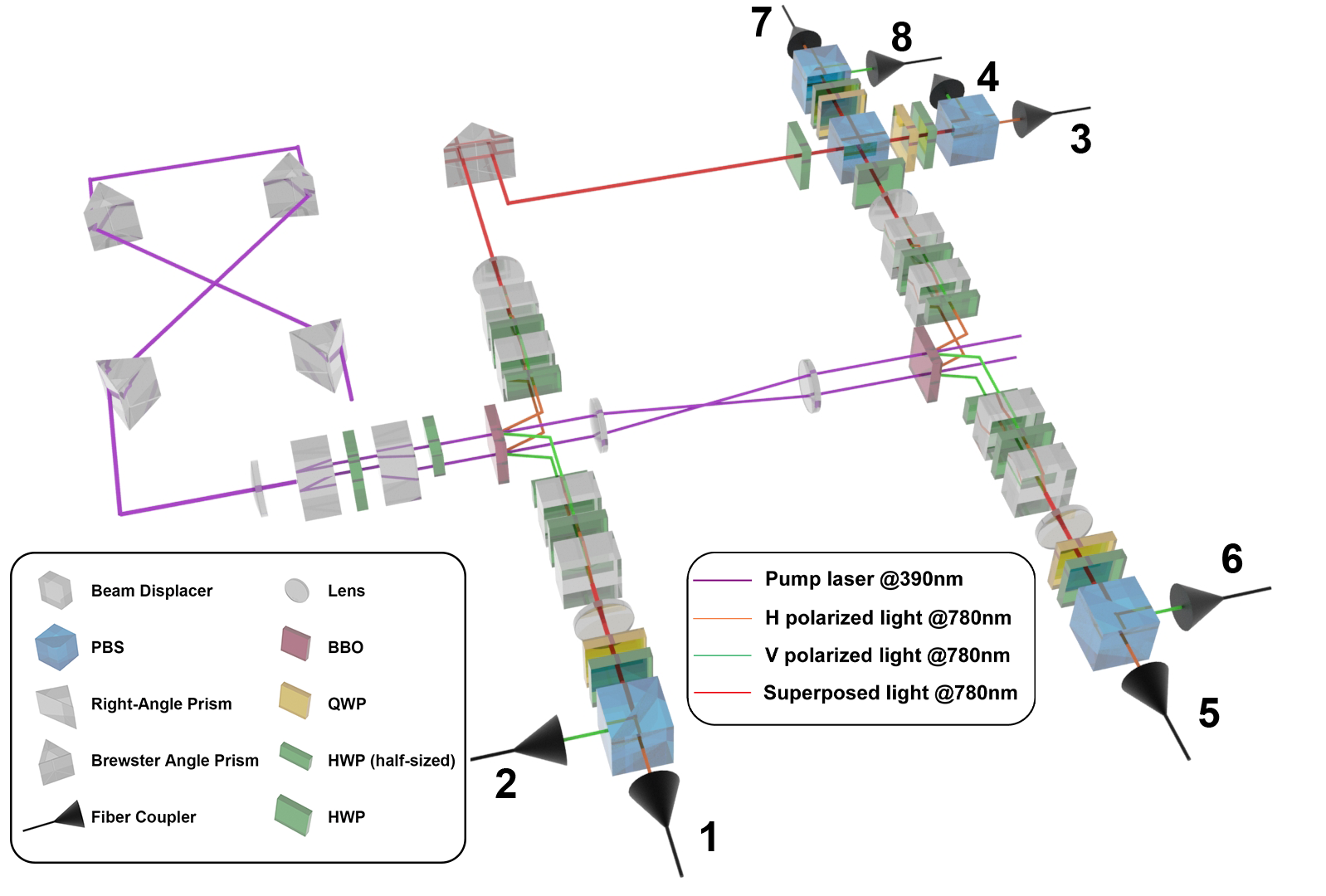}
\caption{Experimental set-up for generating 4-photon entanglement. A Spectra-Physics Tsunami 3960C-15HP ultrafast Ti: Sapphire laser with a central wavelength of 780 nm, the average power of 2.5 W, the repetition rate of 80 MHz, and a pulse width of around 100 fs pumps a Spectra-Physics Inspire Blue femtosecond harmonic generators to generate a near-UV ultrafast pulsed laser with the central wavelength 390 nm, average power around 550 mW, repetition rate 80 MHz and the pulse width around 100 fs. The near-UV pulsed laser is used to pump the SPDC set-up. 4 Brewster prisms are deployed to pre-compensate the mass group delay dispersion of the pulsed laser caused by BDs and BBOs. We use 8 sets of Excelitas SPCM-ARQH-14 single photon counting modules to detect photons. The detectors have a dark count rate of around 100 cps (2‰ of single channel photon count rate), photon detecting efficiency of around 60\%, and dead time of 24 ns. A SIMINICS MT6420 time-correlated single-photon counting system is deployed to perform an 8-channel 16-pattern coincidence counting. The coincidence counting is executed with the firmware of the device and controlled with a PC. The device has a time resolution of 64 ps, the single channel dead time of 50 ns, and the max event rate of 40 M per second. In this experiment, the coincidence counting rate is set to 5 ns. Since the PBS interference is phase insensitive, the observed state can remain stable for days without extra phase stability control.}\label{fig:2}
\end{figure}

Multi-channel coincidence counting is the basic scheme to measure a multi-qubit system. Fig.\ref{fig:2} shows our experimental set-up to generate and measure a 4-qubit GHZ state. We record a 4-photon event when all of the 4 detection units record a photon event within a 5 ns time window. Since there are 2 detectors in each unit, corresponding to 2 different single qubit states, there are 16 possible detector combinations for a recorded 4-photon event corresponding to 16 eigenstates of the measurement basis set (For example, if the measurement basis is set by $\left \{ X,X,X,X \right \} $, the 4-photon event recorded by detector $\left \{ 1,3,5,7 \right \} $ represents the measurement result of 4-qubit state $\ket{DDDD} $). We take counts of all 16 combinations as the raw data to reconstruct the state. In this supplementary material, we note the coincidence count in Table \ref{tab:3}.

\begin{table}[h]
\caption{Notation definition of coincidence count number for the detector combinations}
\begin{tabular}{ c | c | c }
\hline
\hline
Notation & Detector Combination & Corresponding State\\
\hline
$c_1$ & $\left \{ 1,3,5,7 \right \} $ & $\ket{\uparrow\uparrow\uparrow\uparrow}$\\
$c_2$ & $\left \{ 1,3,5,8 \right \} $ & $\ket{\uparrow\uparrow\uparrow\downarrow}$\\
$c_3$ & $\left \{ 1,3,6,7 \right \} $ & $\ket{\uparrow\uparrow\downarrow\uparrow}$\\
$c_4$ & $\left \{ 1,3,6,8 \right \} $ & $\ket{\uparrow\uparrow\downarrow\downarrow}$\\
$c_5$ & $\left \{ 1,4,5,7 \right \} $ & $\ket{\uparrow\downarrow\uparrow\uparrow}$\\
$c_6$ & $\left \{ 1,4,5,8 \right \} $ & $\ket{\uparrow\downarrow\uparrow\downarrow}$\\
$c_7$ & $\left \{ 1,4,6,7 \right \} $ & $\ket{\uparrow\downarrow\downarrow\uparrow}$\\
$c_8$ & $\left \{ 1,4,6,8 \right \} $ & $\ket{\uparrow\downarrow\downarrow\downarrow}$\\
$c_9$ & $\left \{ 2,3,5,7 \right \} $ & $\ket{\downarrow\uparrow\uparrow\uparrow}$\\
$c_{10}$ & $\left \{ 2,3,5,8 \right \} $ & $\ket{\downarrow\uparrow\uparrow\downarrow}$\\
$c_{11}$ & $\left \{ 2,3,6,7 \right \} $ & $\ket{\downarrow\uparrow\downarrow\uparrow}$\\
$c_{12}$ & $\left \{ 2,3,6,8 \right \} $ & $\ket{\downarrow\uparrow\downarrow\downarrow}$\\
$c_{13}$ & $\left \{ 2,4,5,7 \right \} $ & $\ket{\downarrow\downarrow\uparrow\uparrow}$\\
$c_{14}$ & $\left \{ 2,4,5,8 \right \} $ & $\ket{\downarrow\downarrow\uparrow\downarrow}$\\
$c_{15}$ & $\left \{ 2,4,6,7 \right \} $ & $\ket{\downarrow\downarrow\downarrow\uparrow}$\\
$c_{16}$ & $\left \{ 2,4,6,8 \right \} $ & $\ket{\downarrow\downarrow\downarrow\downarrow}$\\
\hline
\hline
\end{tabular}
\label{tab:3} 
\end{table}
\newpage
\subsection{4-qubit state generation}

In our experimental set-up shown in Fig.\ref{fig:2}, we have two SPDC sources generating $\left|\psi_{1,2}\right\rangle=\left|HV\right\rangle+e^{i\theta_1}\left|VH\right.\rangle$ and $\left|\psi_{3,4}\right\rangle=\left|HV\right\rangle+e^{i\theta_2}\left|VH\right.\rangle$.

When inserting HWP a and b into the path and applying rotation to qubit 2 and 4, the states convert into:
\begin{equation}
\left|\psi_{1,2}^\prime\right\rangle=\frac{1}{\sqrt2}(\left(\cos{\theta_a}\left|HV\right\rangle+\sin{\theta_a}\left|HH\right\rangle\right)+e^{i\theta_1}\left(\cos{\theta_a}\left|VH\right\rangle+\sin{\theta_a}\left|VV\right\rangle\right)
\end{equation}
\begin{equation}
\left|\psi_{3,4}^\prime\right\rangle=\frac{1}{\sqrt2}\left(\left(\cos{\theta_b}\left|HV\right\rangle+\sin{\theta_b}\left|HH\right\rangle\right)+e^{i\theta_2}\left(\cos{\theta_b}\left|VH\right\rangle+\sin{\theta_b}\left|VV\right\rangle\right)\right)
\end{equation}

As qubit 2 and 4 interfere at PBS, the wavefunction of the post-selected 4-state is:

\begin{align}
\left|\psi_{1,2,3,4}^\prime\right\rangle
& = \left (  \left | \psi_{1,2}^\prime  \right \rangle \otimes \left | \psi_{3,4}^\prime  \right \rangle \right ) \cdot {\left (  \left | HH  \right \rangle \left \langle HH \right | + \left | VV  \right \rangle \left \langle VV \right | \right )}_{2,4} \notag\\
& = \frac{1}{\sqrt2}(\sin{\theta_a}\sin{\theta_b}\left(\left|HHHH\right\rangle+e^{i\left(\theta_1+\theta_2\right)}\left|VVVV\right\rangle\right)+\sin{\theta_a}\cos{\theta_b}\left(e^{i\theta_2}\left|HHVH\right\rangle-e^{i\theta_1}\left|VVHV\right\rangle\right)\notag\\
& +\cos{\theta_a}\sin{\theta_b}\left(e^{i\theta_1}\left|VHHH\right\rangle-e^{i\theta_2}\left|HVVV\right\rangle\right)+\cos{\theta_a}\cos{\theta_b}\left(\left|HVHV\right\rangle+e^{i\left(\theta_1+\theta_2\right)}\left|VHVH\right\rangle\right))
\end{align}

Inserting $\theta_a=\theta_b=0$ and ${\theta^\prime}_a={\theta^\prime}_b=22.5$, we get $\psi_{GHZ}^R = \frac{1}{\sqrt{2}} (\ket{HVHV} + e^{i\theta}\ket{VHVH})$ (where $\theta=\theta_1+\theta_2$) and ${\psi'}^R=\frac{1}{\sqrt{2}} (\ket{D'VD''V}+\ket{A'HA''H})$ with $\ket{D'}=\ket H+e^{i\theta_1}\ket V, \ket{A'}=\ket H-e^{i\theta_1}\ket V, \ket{D''}=\ket H+e^{i\theta_2}\ket V, \ket{A''}=\ket H-e^{i\theta_2}\ket V$, the two states mentioned in our main text.

\subsection{Frequency correlation and narrow-band filtering}

As we use SPDC to generate polarization entangled photon pairs, we expect the photons are perfectly indistinguishable so that the photon polarization qubits can carry full entanglement information. However, in typical conditions, there would be a strong correlation between the frequencies of the signal and idler photons generated with SPDC \cite{RN27,PhysRevLett.117.210502}. This kind of correlation can be predicted and quantified with the phase-matching equation of SPDC. According to the calculation, the correlation can be considerably eliminated by setting the proper cut angle and thickness of the crystal. But for a given type of crystal and phase matching condition, a total elimination of frequency correlation is only allowed within a specific wavelength window. In a more common condition, we use narrow-band filters to select the photons near the center wavelength, which would have much less frequency correlation.

\begin{figure}[b]
\centering
\includegraphics[width=0.5\linewidth]{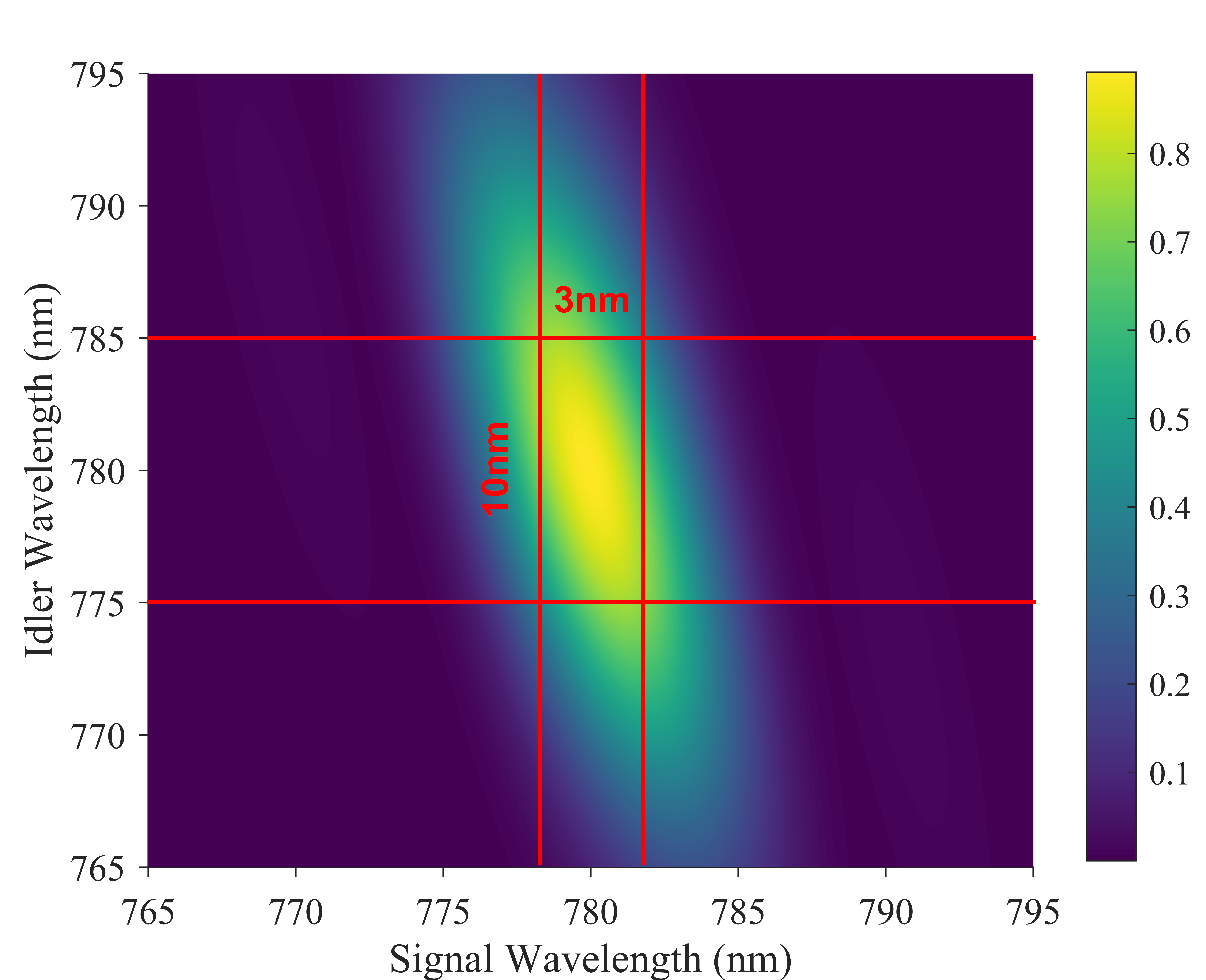}
\caption{The simulated joint spectrum of our SPDC photon pair source. The heatmap describes the relative intensity for the corresponding wavelength combination. The transmission area after narrow-band filtering is highlighted}\label{fig:3}
\end{figure}

Fig.\ref{fig:3} shows a simulation with frequency correlation of the SPDC set-up involved in our experiment. Since there is a significant difference between the spectral peak width of signal and idler photon, it is a natural practice to set filters with different band-width on signal and idler photons respectively. In our case, compromising the brightness, the polarization fidelity, and the availability of the filters, we choose to use a Semrock 780nm maxline\textsuperscript{\textregistered} filter for signal photon and a Thorlabs FBH780-10 filter for idlers photon.

\subsection{Measurement Basis and Data Collection}
Following the scheme of quantum tomography, there are 81 possible basis combinations for the 4-qubit measurement. We measure all of the 81 basis sets for the full tomography. For each basis, we collect the counts for 300s. The total count number of all the 16 coincidence combinations for each measurement basis set in 300s is around 2000 on average.
\subsection{Exact Tomography}
Firstly \cite{Altepeter2005PhotonicST}, we normalize the recorded count numbers $c^{\left \{ B_1, B_2, B_3, B_4 \right \}}_i, i \in \left \{ 1,2, \cdots , 16 \right \}$, $B_j \in \left \{ X, Y, Z \right \}$ representing the measurement basis of qubit j.
\begin{equation}
p^{\left \{ B_1, B_2, B_3, B_4 \right \}}_i = \frac{c^{\left \{ B_1, B_2, B_3, B_4 \right \}}_i }{\sum_{i=1}^{16} c^{\left\{ B_1, B_2, B_3, B_4 \right \}}_i } 
\end{equation}
Following the scheme of quantum tomography, a 4-qubit quantum tomography is to reconstruct a 4-qubit density matrix describing the system:
\begin{equation}
\hat{\rho}=\frac{1}{2^4}\sum_{i_1 ,i_2,i_3 ,i_4 =0}^{3} S_{i_1 ,i_2,i_3 ,i_4} \hat{\sigma}_{i_1}\otimes\hat{\sigma}_{i_2}\otimes\hat{\sigma}_{i_3}\otimes\hat{\sigma}_{i_4},  
\end{equation}
\begin{equation}
\begin{aligned}
i_1 ,i_2,i_3 ,i_4 \in \left \{ 0,1,2,3 \right \}, 
\hat{\sigma}_0\equiv 
\begin{pmatrix}
 1 & 0 \\
 0 & 1 \\
\end{pmatrix}, \notag\\
\hat{\sigma}_1\equiv 
\begin{pmatrix}
 0 & 1\\
 1 & 0
\end{pmatrix}, 
\hat{\sigma}_2\equiv 
\begin{pmatrix}
 0 & -i\\
 i & 0
\end{pmatrix}, 
\hat{\sigma}_3\equiv 
\begin{pmatrix}
 1 & 0\\
 0 & -1
\end{pmatrix}
\end{aligned}
\end{equation}

Noticing that $\hat{\sigma}_{1,2,3}$ are exactly Pauli Matrices $\hat{\sigma}_{x,y,z}$, we can correspond $S_{i_1,i_2,i_3,i_4}$ to $p^{\left \{ B_1, B_2, B_3, B_4 \right \}}_i$ with particular measurement basis. Calculating $S_{i_1,i_2,i_3,i_4} = (P_{\phi_{i_1}}\pm P_{\phi_{i_1}^\bot })(P_{\phi_{i_2}}\pm P_{\phi_{i_2}^\bot })(P_{\phi_{i_3}}\pm P_{\phi_{i_3}^\bot })(P_{\phi_{i_4}}\pm P_{\phi_{i_4}^\bot })$, we have:
\begin{widetext}
\begin{align}
S_{i_1,i_2,i_3,i_4} &= (P_{\phi_{i_1}}- P_{\phi_{i_1}^\bot })(P_{\phi_{i_2}}- P_{\phi_{i_2}^\bot })(P_{\phi_{i_3}}- P_{\phi_{i_3}^\bot })(P_{\phi_{i_4}}- P_{\phi_{i_4}^\bot }) \notag\\
&= P_{\phi_{i_1}}P_{\phi_{i_2}}P_{\phi_{i_3}}P_{\phi_{i_4}} -P_{\phi_{i_1}}P_{\phi_{i_2}}P_{\phi_{i_3}}P_{\phi_{i_4}^\bot} -P_{\phi_{i_1}}P_{\phi_{i_2}}P_{\phi_{i_3}^\bot}P_{\phi_{i_4}}+P_{\phi_{i_1}}P_{\phi_{i_2}}P_{\phi_{i_3}^\bot}P_{\phi_{i_4}^\bot}\notag\\
&-P_{\phi_{i_1}}P_{\phi_{i_2}^\bot}P_{\phi_{i_3}}P_{\phi_{i_4}} +P_{\phi_{i_1}}P_{\phi_{i_2}^\bot}P_{\phi_{i_3}}P_{\phi_{i_4}^\bot} +P_{\phi_{i_1}}P_{\phi_{i_2}^\bot}P_{\phi_{i_3}^\bot}P_{\phi_{i_4}}-P_{\phi_{i_1}}P_{\phi_{i_2}^\bot}P_{\phi_{i_3}^\bot}P_{\phi_{i_4}^\bot}\notag\\
&-P_{\phi_{i_1}^\bot}P_{\phi_{i_2}}P_{\phi_{i_3}}P_{\phi_{i_4}} +P_{\phi_{i_1}^\bot}P_{\phi_{i_2}}P_{\phi_{i_3}}P_{\phi_{i_4}^\bot} +P_{\phi_{i_1}^\bot}P_{\phi_{i_2}}P_{\phi_{i_3}^\bot}P_{\phi_{i_4}}-P_{\phi_{i_1}^\bot}P_{\phi_{i_2}}P_{\phi_{i_3}^\bot}P_{\phi_{i_4}^\bot}\notag\\
&+P_{\phi_{i_1}^\bot}P_{\phi_{i_2}^\bot}P_{\phi_{i_3}}P_{\phi_{i_4}} -P_{\phi_{i_1}^\bot}P_{\phi_{i_2}^\bot}P_{\phi_{i_3}}P_{\phi_{i_4}^\bot} -P_{\phi_{i_1}^\bot}P_{\phi_{i_2}^\bot}P_{\phi_{i_3}^\bot}P_{\phi_{i_4}}+P_{\phi_{i_1}^\bot}P_{\phi_{i_2}^\bot}P_{\phi_{i_3}^\bot}P_{\phi_{i_4}^\bot} \notag\\
&=p_{1}^{\left\{ B_1, B_2, B_3, B_4 \right \}}-p_{2}^{\left\{ B_1, B_2, B_3, B_4 \right \}}-p_{3}^{\left\{ B_1, B_2, B_3, B_4 \right \}}+p_{4}^{\left\{ B_1, B_2, B_3, B_4 \right \}}\notag\\
&-p_{5}^{\left\{ B_1, B_2, B_3, B_4 \right \}}+p_{6}^{\left\{ B_1, B_2, B_3, B_4 \right \}}+p_{7}^{\left\{ B_1, B_2, B_3, B_4 \right \}}-p_{8}^{\left\{ B_1, B_2, B_3, B_4 \right \}}\notag\\
&-p_{9}^{\left\{ B_1, B_2, B_3, B_4 \right \}}+p_{10}^{\left\{ B_1, B_2, B_3, B_4 \right \}}+p_{11}^{\left\{ B_1, B_2, B_3, B_4 \right \}}-p_{12}^{\left\{ B_1, B_2, B_3, B_4 \right \}}\notag\\
&+p_{13}^{\left\{ B_1, B_2, B_3, B_4 \right \}}-p_{14}^{\left\{ B_1, B_2, B_3, B_4 \right \}}-p_{15}^{\left\{ B_1, B_2, B_3, B_4 \right \}}+p_{16}^{\left\{ B_1, B_2, B_3, B_4 \right \}}
\end{align}
\end{widetext}
Here, $i_1,i_2,i_3,i_4 = 1,2,3$, $\left\{ B_1, B_2, B_3, B_4 \right \}$ is the corresponding measurement basis set to $\left\{ i_1,i_2,i_3,i_4 \right \}$. For example, for $i_1=i_2=i_3=i_4=1$,  $\left\{ B_1, B_2, B_3, B_4 \right \}$ would be $\left\{ X,X,X,X \right \}$. Starting with the same base, we also can derive the S value with $i=0$:

\begin{widetext}

\begin{align}
S_{0,i_2,i_3,i_4} &= (P_{\phi_{i_2}}- P_{\phi_{i_2}^\bot })(P_{\phi_{i_3}}- P_{\phi_{i_3}^\bot })(P_{\phi_{i_4}}- P_{\phi_{i_4}^\bot }) \notag\\
&=p_{1}^{\left\{ B_1, B_2, B_3, B_4 \right \}}-p_{2}^{\left\{ B_1, B_2, B_3, B_4 \right \}}-p_{3}^{\left\{ B_1, B_2, B_3, B_4 \right \}}+p_{4}^{\left\{ B_1, B_2, B_3, B_4 \right \}}\notag\\
&-p_{5}^{\left\{ B_1, B_2, B_3, B_4 \right \}}+p_{6}^{\left\{ B_1, B_2, B_3, B_4 \right \}}+p_{7}^{\left\{ B_1, B_2, B_3, B_4 \right \}}-p_{8}^{\left\{ B_1, B_2, B_3, B_4 \right \}}\notag\\
&+p_{9}^{\left\{ B_1, B_2, B_3, B_4 \right \}}-p_{10}^{\left\{ B_1, B_2, B_3, B_4 \right \}}-p_{11}^{\left\{ B_1, B_2, B_3, B_4 \right \}}+p_{12}^{\left\{ B_1, B_2, B_3, B_4 \right \}}\notag\\
&-p_{13}^{\left\{ B_1, B_2, B_3, B_4 \right \}}+p_{14}^{\left\{ B_1, B_2, B_3, B_4 \right \}}+p_{15}^{\left\{ B_1, B_2, B_3, B_4 \right \}}-p_{16}^{\left\{ B_1, B_2, B_3, B_4 \right \}}
\end{align}

\begin{align}
S_{i_1,0,i_3,i_4}
& = p_{1}^{\left\{ B_1, B_2, B_3, B_4 \right \}} - p_{2}^{\left\{ B_1, B_2, B_3, B_4 \right \}} - p_{3}^{\left\{ B_1, B_2, B_3, B_4 \right \}} + p_{4}^{\left\{ B_1, B_2, B_3, B_4 \right \}}\notag\\
& + p_{5}^{\left\{ B_1, B_2, B_3, B_4 \right \}} - p_{6}^{\left\{ B_1, B_2, B_3, B_4 \right \}} - p_{7}^{\left\{ B_1, B_2, B_3, B_4 \right \}} + p_{8}^{\left\{ B_1, B_2, B_3, B_4 \right \}}\notag\\
& - p_{9}^{\left\{ B_1, B_2, B_3, B_4 \right \}} + p_{10}^{\left\{ B_1, B_2, B_3, B_4 \right \}} + p_{11}^{\left\{ B_1, B_2, B_3, B_4 \right \}} - p_{12}^{\left\{ B_1, B_2, B_3, B_4 \right \}}\notag\\
& - p_{13}^{\left\{ B_1, B_2, B_3, B_4 \right \}} + p_{14}^{\left\{ B_1, B_2, B_3, B_4 \right \}} + p_{15}^{\left\{ B_1, B_2, B_3, B_4 \right \}} - p_{16}^{\left\{ B_1, B_2, B_3, B_4 \right \}}
\end{align}

\begin{align}
S_{i_1,i_2,0,i_4}
& = p_{1}^{\left\{ B_1, B_2, B_3, B_4 \right \}} - p_{2}^{\left\{ B_1, B_2, B_3, B_4 \right \}} + p_{3}^{\left\{ B_1, B_2, B_3, B_4 \right \}} - p_{4}^{\left\{ B_1, B_2, B_3, B_4 \right \}}\notag\\
& - p_{5}^{\left\{ B_1, B_2, B_3, B_4 \right \}} + p_{6}^{\left\{ B_1, B_2, B_3, B_4 \right \}} - p_{7}^{\left\{ B_1, B_2, B_3, B_4 \right \}} + p_{8}^{\left\{ B_1, B_2, B_3, B_4 \right \}}\notag\\
& - p_{9}^{\left\{ B_1, B_2, B_3, B_4 \right \}} + p_{10}^{\left\{ B_1, B_2, B_3, B_4 \right \}} - p_{11}^{\left\{ B_1, B_2, B_3, B_4 \right \}} + p_{12}^{\left\{ B_1, B_2, B_3, B_4 \right \}}\notag\\
& + p_{13}^{\left\{ B_1, B_2, B_3, B_4 \right \}} - p_{14}^{\left\{ B_1, B_2, B_3, B_4 \right \}} + p_{15}^{\left\{ B_1, B_2, B_3, B_4 \right \}} - p_{16}^{\left\{ B_1, B_2, B_3, B_4 \right \}}
\end{align}

\begin{align}
S_{i_1,i_2,i_3,0}
& = p_{1}^{\left\{ B_1, B_2, B_3, B_4 \right \}} + p_{2}^{\left\{ B_1, B_2, B_3, B_4 \right \}} - p_{3}^{\left\{ B_1, B_2, B_3, B_4 \right \}} - p_{4}^{\left\{ B_1, B_2, B_3, B_4 \right \}}\notag\\
& - p_{5}^{\left\{ B_1, B_2, B_3, B_4 \right \}} - p_{6}^{\left\{ B_1, B_2, B_3, B_4 \right \}} + p_{7}^{\left\{ B_1, B_2, B_3, B_4 \right \}} + p_{8}^{\left\{ B_1, B_2, B_3, B_4 \right \}}\notag\\
& - p_{9}^{\left\{ B_1, B_2, B_3, B_4 \right \}} - p_{10}^{\left\{ B_1, B_2, B_3, B_4 \right \}} + p_{11}^{\left\{ B_1, B_2, B_3, B_4 \right \}} + p_{12}^{\left\{ B_1, B_2, B_3, B_4 \right \}}\notag\\
& + p_{13}^{\left\{ B_1, B_2, B_3, B_4 \right \}} + p_{14}^{\left\{ B_1, B_2, B_3, B_4 \right \}} - p_{15}^{\left\{ B_1, B_2, B_3, B_4 \right \}} - p_{16}^{\left\{ B_1, B_2, B_3, B_4 \right \}}
\end{align}

\begin{align}
S_{0,0,i_3,i_4}
& = p_{1}^{\left\{ B_1, B_2, B_3, B_4 \right \}} - p_{2}^{\left\{ B_1, B_2, B_3, B_4 \right \}} - p_{3}^{\left\{ B_1, B_2, B_3, B_4 \right \}} + p_{4}^{\left\{ B_1, B_2, B_3, B_4 \right \}}\notag\\
& + p_{5}^{\left\{ B_1, B_2, B_3, B_4 \right \}} - p_{6}^{\left\{ B_1, B_2, B_3, B_4 \right \}} - p_{7}^{\left\{ B_1, B_2, B_3, B_4 \right \}} + p_{8}^{\left\{ B_1, B_2, B_3, B_4 \right \}}\notag\\
& + p_{9}^{\left\{ B_1, B_2, B_3, B_4 \right \}} - p_{10}^{\left\{ B_1, B_2, B_3, B_4 \right \}} - p_{11}^{\left\{ B_1, B_2, B_3, B_4 \right \}} + p_{12}^{\left\{ B_1, B_2, B_3, B_4 \right \}}\notag\\
& + p_{13}^{\left\{ B_1, B_2, B_3, B_4 \right \}} - p_{14}^{\left\{ B_1, B_2, B_3, B_4 \right \}} - p_{15}^{\left\{ B_1, B_2, B_3, B_4 \right \}} + p_{16}^{\left\{ B_1, B_2, B_3, B_4 \right \}}
\end{align}

\begin{align}
S_{0,i_2,0,i_4}
& = p_{1}^{\left\{ B_1, B_2, B_3, B_4 \right \}} - p_{2}^{\left\{ B_1, B_2, B_3, B_4 \right \}} + p_{3}^{\left\{ B_1, B_2, B_3, B_4 \right \}} - p_{4}^{\left\{ B_1, B_2, B_3, B_4 \right \}}\notag\\
& - p_{5}^{\left\{ B_1, B_2, B_3, B_4 \right \}} + p_{6}^{\left\{ B_1, B_2, B_3, B_4 \right \}} - p_{7}^{\left\{ B_1, B_2, B_3, B_4 \right \}} + p_{8}^{\left\{ B_1, B_2, B_3, B_4 \right \}}\notag\\
& + p_{9}^{\left\{ B_1, B_2, B_3, B_4 \right \}} - p_{10}^{\left\{ B_1, B_2, B_3, B_4 \right \}} + p_{11}^{\left\{ B_1, B_2, B_3, B_4 \right \}} - p_{12}^{\left\{ B_1, B_2, B_3, B_4 \right \}}\notag\\
& - p_{13}^{\left\{ B_1, B_2, B_3, B_4 \right \}} + p_{14}^{\left\{ B_1, B_2, B_3, B_4 \right \}} - p_{15}^{\left\{ B_1, B_2, B_3, B_4 \right \}} + p_{16}^{\left\{ B_1, B_2, B_3, B_4 \right \}}
\end{align}

\begin{align}
S_{0,i_2,i_3,0}
& = p_{1}^{\left\{ B_1, B_2, B_3, B_4 \right \}} + p_{2}^{\left\{ B_1, B_2, B_3, B_4 \right \}} - p_{3}^{\left\{ B_1, B_2, B_3, B_4 \right \}} - p_{4}^{\left\{ B_1, B_2, B_3, B_4 \right \}}\notag\\
& - p_{5}^{\left\{ B_1, B_2, B_3, B_4 \right \}} - p_{6}^{\left\{ B_1, B_2, B_3, B_4 \right \}} + p_{7}^{\left\{ B_1, B_2, B_3, B_4 \right \}} + p_{8}^{\left\{ B_1, B_2, B_3, B_4 \right \}}\notag\\
& + p_{9}^{\left\{ B_1, B_2, B_3, B_4 \right \}} + p_{10}^{\left\{ B_1, B_2, B_3, B_4 \right \}} - p_{11}^{\left\{ B_1, B_2, B_3, B_4 \right \}} - p_{12}^{\left\{ B_1, B_2, B_3, B_4 \right \}}\notag\\
& - p_{13}^{\left\{ B_1, B_2, B_3, B_4 \right \}} - p_{14}^{\left\{ B_1, B_2, B_3, B_4 \right \}} + p_{15}^{\left\{ B_1, B_2, B_3, B_4 \right \}} + p_{16}^{\left\{ B_1, B_2, B_3, B_4 \right \}}
\end{align}

\begin{align}
S_{i_1,0,0,i_4}
& = p_{1}^{\left\{ B_1, B_2, B_3, B_4 \right \}} - p_{2}^{\left\{ B_1, B_2, B_3, B_4 \right \}} + p_{3}^{\left\{ B_1, B_2, B_3, B_4 \right \}} - p_{4}^{\left\{ B_1, B_2, B_3, B_4 \right \}}\notag\\
& + p_{5}^{\left\{ B_1, B_2, B_3, B_4 \right \}} - p_{6}^{\left\{ B_1, B_2, B_3, B_4 \right \}} + p_{7}^{\left\{ B_1, B_2, B_3, B_4 \right \}} - p_{8}^{\left\{ B_1, B_2, B_3, B_4 \right \}}\notag\\
& - p_{9}^{\left\{ B_1, B_2, B_3, B_4 \right \}} + p_{10}^{\left\{ B_1, B_2, B_3, B_4 \right \}} - p_{11}^{\left\{ B_1, B_2, B_3, B_4 \right \}} + p_{12}^{\left\{ B_1, B_2, B_3, B_4 \right \}}\notag\\
& - p_{13}^{\left\{ B_1, B_2, B_3, B_4 \right \}} + p_{14}^{\left\{ B_1, B_2, B_3, B_4 \right \}} - p_{15}^{\left\{ B_1, B_2, B_3, B_4 \right \}} + p_{16}^{\left\{ B_1, B_2, B_3, B_4 \right \}}
\end{align}

\begin{align}
S_{i_1,0,i_3,0}
& = p_{1}^{\left\{ B_1, B_2, B_3, B_4 \right \}} + p_{2}^{\left\{ B_1, B_2, B_3, B_4 \right \}} - p_{3}^{\left\{ B_1, B_2, B_3, B_4 \right \}} - p_{4}^{\left\{ B_1, B_2, B_3, B_4 \right \}}\notag\\
& + p_{5}^{\left\{ B_1, B_2, B_3, B_4 \right \}} + p_{6}^{\left\{ B_1, B_2, B_3, B_4 \right \}} - p_{7}^{\left\{ B_1, B_2, B_3, B_4 \right \}} - p_{8}^{\left\{ B_1, B_2, B_3, B_4 \right \}}\notag\\
& - p_{9}^{\left\{ B_1, B_2, B_3, B_4 \right \}} - p_{10}^{\left\{ B_1, B_2, B_3, B_4 \right \}} + p_{11}^{\left\{ B_1, B_2, B_3, B_4 \right \}} + p_{12}^{\left\{ B_1, B_2, B_3, B_4 \right \}}\notag\\
& - p_{13}^{\left\{ B_1, B_2, B_3, B_4 \right \}} - p_{14}^{\left\{ B_1, B_2, B_3, B_4 \right \}} + p_{15}^{\left\{ B_1, B_2, B_3, B_4 \right \}} + p_{16}^{\left\{ B_1, B_2, B_3, B_4 \right \}}
\end{align}

\begin{align}
S_{i_1,i_2,0,0}
& = p_{1}^{\left\{ B_1, B_2, B_3, B_4 \right \}} + p_{2}^{\left\{ B_1, B_2, B_3, B_4 \right \}} + p_{3}^{\left\{ B_1, B_2, B_3, B_4 \right \}} + p_{4}^{\left\{ B_1, B_2, B_3, B_4 \right \}}\notag\\
& - p_{5}^{\left\{ B_1, B_2, B_3, B_4 \right \}} - p_{6}^{\left\{ B_1, B_2, B_3, B_4 \right \}} - p_{7}^{\left\{ B_1, B_2, B_3, B_4 \right \}} - p_{8}^{\left\{ B_1, B_2, B_3, B_4 \right \}}\notag\\
& - p_{9}^{\left\{ B_1, B_2, B_3, B_4 \right \}} - p_{10}^{\left\{ B_1, B_2, B_3, B_4 \right \}} - p_{11}^{\left\{ B_1, B_2, B_3, B_4 \right \}} - p_{12}^{\left\{ B_1, B_2, B_3, B_4 \right \}}\notag\\
& + p_{13}^{\left\{ B_1, B_2, B_3, B_4 \right \}} + p_{14}^{\left\{ B_1, B_2, B_3, B_4 \right \}} + p_{15}^{\left\{ B_1, B_2, B_3, B_4 \right \}} + p_{16}^{\left\{ B_1, B_2, B_3, B_4 \right \}}
\end{align}

\begin{align}
S_{i_1,0,0,0}
& = p_{1}^{\left\{ B_1, B_2, B_3, B_4 \right \}} + p_{2}^{\left\{ B_1, B_2, B_3, B_4 \right \}} + p_{3}^{\left\{ B_1, B_2, B_3, B_4 \right \}} + p_{4}^{\left\{ B_1, B_2, B_3, B_4 \right \}}\notag\\
& + p_{5}^{\left\{ B_1, B_2, B_3, B_4 \right \}} + p_{6}^{\left\{ B_1, B_2, B_3, B_4 \right \}} + p_{7}^{\left\{ B_1, B_2, B_3, B_4 \right \}} + p_{8}^{\left\{ B_1, B_2, B_3, B_4 \right \}}\notag\\
& - p_{9}^{\left\{ B_1, B_2, B_3, B_4 \right \}} - p_{10}^{\left\{ B_1, B_2, B_3, B_4 \right \}} - p_{11}^{\left\{ B_1, B_2, B_3, B_4 \right \}} - p_{12}^{\left\{ B_1, B_2, B_3, B_4 \right \}}\notag\\
& - p_{13}^{\left\{ B_1, B_2, B_3, B_4 \right \}} - p_{14}^{\left\{ B_1, B_2, B_3, B_4 \right \}} - p_{15}^{\left\{ B_1, B_2, B_3, B_4 \right \}} - p_{16}^{\left\{ B_1, B_2, B_3, B_4 \right \}}
\end{align}

\begin{align}
S_{0,i_2,0,0}
& = p_{1}^{\left\{ B_1, B_2, B_3, B_4 \right \}} + p_{2}^{\left\{ B_1, B_2, B_3, B_4 \right \}} + p_{3}^{\left\{ B_1, B_2, B_3, B_4 \right \}} + p_{4}^{\left\{ B_1, B_2, B_3, B_4 \right \}}\notag\\
& - p_{5}^{\left\{ B_1, B_2, B_3, B_4 \right \}} - p_{6}^{\left\{ B_1, B_2, B_3, B_4 \right \}} - p_{7}^{\left\{ B_1, B_2, B_3, B_4 \right \}} - p_{8}^{\left\{ B_1, B_2, B_3, B_4 \right \}}\notag\\
& + p_{9}^{\left\{ B_1, B_2, B_3, B_4 \right \}} + p_{10}^{\left\{ B_1, B_2, B_3, B_4 \right \}} + p_{11}^{\left\{ B_1, B_2, B_3, B_4 \right \}} + p_{12}^{\left\{ B_1, B_2, B_3, B_4 \right \}}\notag\\
& - p_{13}^{\left\{ B_1, B_2, B_3, B_4 \right \}} - p_{14}^{\left\{ B_1, B_2, B_3, B_4 \right \}} - p_{15}^{\left\{ B_1, B_2, B_3, B_4 \right \}} - p_{16}^{\left\{ B_1, B_2, B_3, B_4 \right \}}
\end{align}

\begin{align}
S_{0,0,i_3,0}
& = p_{1}^{\left\{ B_1, B_2, B_3, B_4 \right \}} + p_{2}^{\left\{ B_1, B_2, B_3, B_4 \right \}} - p_{3}^{\left\{ B_1, B_2, B_3, B_4 \right \}} - p_{4}^{\left\{ B_1, B_2, B_3, B_4 \right \}}\notag\\
& + p_{5}^{\left\{ B_1, B_2, B_3, B_4 \right \}} + p_{6}^{\left\{ B_1, B_2, B_3, B_4 \right \}} - p_{7}^{\left\{ B_1, B_2, B_3, B_4 \right \}} - p_{8}^{\left\{ B_1, B_2, B_3, B_4 \right \}}\notag\\
& + p_{9}^{\left\{ B_1, B_2, B_3, B_4 \right \}} + p_{10}^{\left\{ B_1, B_2, B_3, B_4 \right \}} - p_{11}^{\left\{ B_1, B_2, B_3, B_4 \right \}} - p_{12}^{\left\{ B_1, B_2, B_3, B_4 \right \}}\notag\\
& + p_{13}^{\left\{ B_1, B_2, B_3, B_4 \right \}} + p_{14}^{\left\{ B_1, B_2, B_3, B_4 \right \}} - p_{15}^{\left\{ B_1, B_2, B_3, B_4 \right \}} - p_{16}^{\left\{ B_1, B_2, B_3, B_4 \right \}}
\end{align}

\begin{align}
S_{0,0,0,i_4}
& = p_{1}^{\left\{ B_1, B_2, B_3, B_4 \right \}} - p_{2}^{\left\{ B_1, B_2, B_3, B_4 \right \}} + p_{3}^{\left\{ B_1, B_2, B_3, B_4 \right \}} - p_{4}^{\left\{ B_1, B_2, B_3, B_4 \right \}}\notag\\
& + p_{5}^{\left\{ B_1, B_2, B_3, B_4 \right \}} - p_{6}^{\left\{ B_1, B_2, B_3, B_4 \right \}} + p_{7}^{\left\{ B_1, B_2, B_3, B_4 \right \}} - p_{8}^{\left\{ B_1, B_2, B_3, B_4 \right \}}\notag\\
& + p_{9}^{\left\{ B_1, B_2, B_3, B_4 \right \}} - p_{10}^{\left\{ B_1, B_2, B_3, B_4 \right \}} + p_{11}^{\left\{ B_1, B_2, B_3, B_4 \right \}} - p_{12}^{\left\{ B_1, B_2, B_3, B_4 \right \}}\notag\\
& + p_{13}^{\left\{ B_1, B_2, B_3, B_4 \right \}} - p_{14}^{\left\{ B_1, B_2, B_3, B_4 \right \}} + p_{15}^{\left\{ B_1, B_2, B_3, B_4 \right \}} - p_{16}^{\left\{ B_1, B_2, B_3, B_4 \right \}}
\end{align}

\begin{align}
S_{0,0,0,0} = 1
\end{align}

\end{widetext}

Then for each term $S_{i_1,i_2,i_3,i_4}$ we can calculate the value with the measurement result from a particular basis $\left \{ B_1, B_2, B_3, B_4 \right \}$. Noticing that for $\left \{ i_1,i_2,i_3,i_4 \right \}$ with any $i=0$, for example, $S_{0,i_2,i_3,i_4}$, there is more than one basis corresponding to this value. In this case, we have $\left \{X, B_2, B_3, B_4 \right \}$, $\left \{ Y, B_2, B_3, B_4 \right \}$ and $\left \{ Z, B_2, B_3, B_4 \right \}$. Theoretically they should give the same value. In our experiment, we simply count the average value to suppress the systematic inaccuracy.

\begin{equation}
S_{0,i_2,i_3,i_4} = \frac{S_{0,i_2,i_3,i_4}^{\left \{X, B_2, B_3, B_4\right \}} + S_{0,i_2,i_3,i_4}^{\left \{ Y, B_2, B_3, B_4\right \}} + S_{0,i_2,i_3,i_4}^{\left \{ Z, B_2, B_3, B_4\right \}}}{3}
\end{equation}

So far we've calculated all $S_{i_1,i_2,i_3,i_4}$ in Eq(2), which means the full density matrix of the 4-qubit system can be reconstructed. In our experiment, we operated a full tomography on the 4-qubit GHZ state to identify the exact wavefunction and the state fidelity.

\subsection{Bayesian State Estimation}
According to Bayes' theorem, a conditional probability can be calculated by
\begin{equation}
P(A|B)=\frac{P(B|A)P(A)}{P(B)}.
\end{equation}

For quantum state estimation \cite{Blume_Kohout_2010}, A represents for the parameter set $\vec{\theta} = \left\{ \theta_1, \theta_2, \cdots , \theta_D\right\}$ that determines density matrix $\rho(\vec{\theta})$ , B represents the measurement dataset $\vec{X}$. In a single quantum measurement, only one set of $\vec{X}$ is gathered, $P(\vec{X})=1$. $P(\vec{X}|\vec{\theta})$ defines likelihood $L_{\vec{X}} (\vec{\theta})$. Then the probability distribution of parameter set $\vec{\theta}$ can be expressed as

\begin{equation}
\pi(\vec{\theta})=L_{\vec{X}} (\vec{\theta})\pi_0(\vec{\theta})
\end{equation}

Here, $\pi(\vec{\theta})$ is the estimated probability distribution of $\vec{\theta}$, and $\pi_0(\vec{\theta})$ is the prior distribution representing a belief about the result before the estimation \cite{Lukens_2020}.

In our case, firstly we express likelihood as

\begin{equation}
L_{\vec{X}}(\vec{\theta}) = \prod \exp \left( -\frac{(\overline{n}_{i,j}(\vec{\theta})-n_{i,j})^2}{\overline{n}_{i,j}(\vec{\theta})} \right)
\end{equation}

Here $\left\{ n_{i,j} \right\}$ stands for the measurement result dataset, and $\left\{ \overline{n}_{i,j}(\vec{\theta}) \right\}$ stands for the expectation measurement result dataset corresponding to the density matrix $\rho(\vec{\theta})$ determined by parameter set $\vec{\theta}$.

Then we discuss the prior distribution in this case. Technically the prior distribution should be a uniform distribution as we have no idea what the state is before the measurement. But in our case, as we have measured and calculated with exact tomography on the set-up, we have the expectation with the state. Thus we use a prior distribution as

\begin{equation}
\pi_0(\vec{\theta}) = \prod \exp \left( -\frac{(\theta_i-\theta_i^R)^2}{2\theta_i^R} \right)
\end{equation}

Here $\vec{\theta}^R = \left\{ \theta_1^R, \theta_2^R, \cdots , \theta_N^R\right\}$ stands for the parameter set corresponding to a reference state $\psi^R$, which is the target state  we should have generated with perfect experimental set-up and we have used to calculate quantum state fidelity with our result. The state is estimated based on both theoretical inference and the result of exact tomography.

With access to $\pi(\vec{\theta})$, we can obtain the expectation value of the density matrix $\rho$

\begin{equation}
\langle \rho \rangle = \int d\vec{\theta} \pi(\vec{\theta})\rho(\vec{\theta})
\end{equation}

\subsection{Gibbs Sampling}

Though Eq(24) determines an estimation of density matrix $\rho$, it is difficult to calculate the integral directly. In computer science, the common approach to solving this kind of question is to obtain a set of random samples $\left \{ \vec{\theta}^{(1)}, \vec{\theta}^{(2)}, \cdots, \vec{\theta}^{(R)} \right \}$ following the target distribution $\pi(\vec{\theta})$. As the sample size R goes considerably large, Eq(24) can be approximated as

\begin{equation}
\langle \rho \rangle \approx \frac{1}{R} \sum_{r=1}^R\rho(\vec{\theta}^{(r)})
\end{equation}

To determine a 4-qubit density matrix, a dataset with 255 parameters is required. In our experiment, we choose $\{S_{i_1 ,i_2,i_3 ,i_4}\}$ in Eq(2) as dataset. To perform sampling over a dataset with such a high dimension, we use Gibbs sampling algorithm\cite{4767596}, a branch of the Markov chain Monte Carlo (MCMC) method. The workflow is as below:

\begin{itemize}
\item[a)] Set step size $\beta$. Set $j=0, k=1$. Use exact tomography result $\vec{\theta}^E$ as starting point $\vec{\theta}^{(0)}_{(1)}=\{\theta_1^{(0)}, \theta_2^{(0)}, \cdots, \theta_D^{(0)}\}$
\item[b)] Propose new parameter $\theta_k'$ by generating a random number following normal distribution $N(\theta_k^{(j)},\beta\theta_k^{(j)})$, noting $\vec{\theta}'=\{\theta_1^{(j+1)}, \theta_2^{(j+1)}, \cdots,\theta_k',\cdots, \theta_D^{(j)}\}$
\item[c)] Calculate acceptance probability by $A(\vec{\theta}',\vec{\theta}_{(k)}^{(j)})$ following
\begin{equation}
\log A(\vec{\theta}',\vec{\theta}_{(k)}^{(j)}) = \min\left \{0, \left(\log L_{\vec{X}}(\vec{\theta}') +\log \pi_0(\vec{\theta}')\right) -\left(\log L_{\vec{X}}(\vec{\theta}_{(k)}^{(j)} +\log \pi_0(\vec{\theta}_{(k)}^{(j)}))\right) - \left(\log \frac{g(\theta_k'|\theta_k^{(j)})}{g(\theta_k^{(j)}|\theta_k')}\right) \right \}
\end{equation}
Here proposal distribution $g(\theta_k'|\theta_k^{(j)})$ is probability of proposing $\theta_k'$ with given $\theta_k^{(j)}$ in step b. In this case the probability is described with a normal distribution.
\item[d)] Set $\theta_k^{(j+1)}=\theta_k'$ with probability $A(\vec{\theta}',\vec{\theta}_{(k)}^{(j)})$, otherwise $\theta_k^{(j+1)}=\theta_k^{(j)}$, noting $\vec{\theta}_{(k+1)}^{(j)}=\{\theta_1^{(j+1)}, \theta_2^{(j+1)}, \cdots,\theta_k^{(j+1)},$ $\cdots, \theta_D^{(j)}\}$, increment $k$ by 1.
\item[e)] If $k$ reaches N+1, set $k$ to 1 and increment $j$ by 1. Return to step b.
\end{itemize}

\subsection{Discussion}

In this experiment, the 4-qubit quantum fidelity is considerably low for both two states we generated, especially for $\psi'$ with a 4-fold fidelity of only 0.836. Several factors contribute significantly to this situation. 

Firstly, the 2-fold fidelity of the SPDC photon pair is not the best. There are three major factors here. First, we pump the SPDC crystal with a high laser power at 550 mW, which can cause a significant multi-photon emission in the SPDC and lower the measurement fidelity. Second, the BD recombination of one photon pair was not good during this experiment. There is a significant temporal mismatch in one of the channels, which causes the fidelity of one SPDC pair to be measured significantly lower than the other in the pre-experiment test, leading to a contribution to the 4-fold infidelity. Third, limited to the availability, we choose a narrow-band filter with 10nm bandwidth on the idler channel, instead of one with 8nm bandwidth which is more widely accepted in similar cases. A wider bandpass offers better brightness but worse fidelity due to the increased frequency correlation.

Besides, we use all broadband PBSs with $T_p$ around 92\% and $R_s$ around 88\% in our experiment. Although the extinction rate is reasonably high over 500:1 (still not good), the low $T_p$ and $R_s$, and the significant unbalance between them, lead to low and unbalanced collecting efficiencies for all the channels, as well as a non-perfect PBS interference. The PBS factors also contribute to infidelity.

Finally, though the SPDC set-up can remain stable for weeks in principle, in practice random environment factors such as unstable temperature and room vibration can cause slight shifting of the light-path, which can lead to a significant influence on the set-up by affecting the photon collection and PBS interference. There is about one week between we generate $\psi$ and $\psi'$. The stability of the set-up when we finally measure $\psi'$ is significantly worse. We believe this could explain the difference between the fidelities of the two states.

\section{Experimental Methodology of Overlapping Tomography}

Following the scheme of QOT, we divide the 4 qubit system, noting as $\left \{1,2,3,4\right \}$, in 2 different ways, $\left \{\left \{1,2\right \}, \left \{3,4\right \}\right \}$ and $\left \{\left \{1,3\right \}, \left \{2,4\right \}\right \}$. 15 measurement basis sets in total are necessary for this tomography, including $\left \{X,X,Y,Y\right \}$, $\left \{X,X,Z,Z\right \}$, $\left \{Y,Y,X,X\right \}$, $\left \{Y,Y,Z,Z\right \}$, $\left \{Z,Z,X,X\right \}$, $\left \{Z,Z,Y,Y\right \}$ to demonstrate $\left \{\left \{1,2\right \}, \left \{3,4\right \}\right \}$, $\left \{X,Y,X,Y\right \}$, $\left \{X,Z,X,Z\right \}$, $\left \{Y,X,Y,X\right \}$, $\left \{Y,Z,Y,Z\right \}$, $\left \{Z,X,Z,X\right \}$, $\left \{Z,Y,Z,Y\right \}$ for $\left \{\left \{1,3\right \}, \left \{2,4\right \}\right \}$, and $\left \{X,X,X,X\right \}$, $\left \{Y,Y,Y,Y\right \}$, $\left \{Z,Z,Z,Z\right \}$

Noticing that to reconstruct density matrices of the 2-qubit subsystems

\begin{equation}
\hat{\rho}^{\left \{ x_1,x_2 \right \} } = \frac{1}{2^2} \sum_{i_1,i_2=0}^{3} S_{i_1,i_2}^{\left \{ x_1,x_2 \right \} }\hat{\sigma}_{i_1}\otimes \hat{\sigma}_{i_2},
\end{equation}

we need to extract $S_{i_1,i_2}^{\left \{ x_1,x_2 \right \} }$. Now we calculate these S values following the rules below:

\begin{itemize}

\item[a.]
For all $S_{i_1,i_2}^{\left \{ x_1,x_2 \right \} }$ with $i_1=i_2=i$, use the measurement result with basis $\left \{X,X,X,X\right \}$, $\left \{Y,Y,Y,Y\right \}$, $\left \{Z,Z,Z,Z\right \}$. We calculate the 2-fold terms following Eq(8)-(13) and apply these values to corresponding S, For example, 
\begin{widetext}
\begin{align}
S_{1,1}^{\left \{ 2,4 \right \} } = S_{0,1,0,1}
& = p_{1}^{\left\{ X,X,X,X \right \}} - p_{2}^{\left\{ X,X,X,X \right \}} + p_{3}^{\left\{ X,X,X,X \right \}} - p_{4}^{\left\{ X,X,X,X \right \}}\notag\\
& - p_{5}^{\left\{ X,X,X,X \right \}} + p_{6}^{\left\{ X,X,X,X \right \}} - p_{7}^{\left\{ X,X,X,X \right \}} + p_{8}^{\left\{ X,X,X,X \right \}}\notag\\
& + p_{9}^{\left\{ X,X,X,X \right \}} - p_{10}^{\left\{ X,X,X,X \right \}} + p_{11}^{\left\{ X,X,X,X \right \}} - p_{12}^{\left\{ X,X,X,X \right \}}\notag\\
& - p_{13}^{\left\{ X,X,X,X \right \}} + p_{14}^{\left\{ X,X,X,X \right \}} - p_{15}^{\left\{X,X,X,X \right \}} + p_{16}^{\left\{ X,X,X,X \right \}}
\end{align}
\end{widetext}
\item[b.]
For $S_{i_1,i_2}^{\left \{ x_1,x_2 \right \} }$ with $i_1 \ne i_2$, $\left \{ x_1,x_2 \right \} = \left \{ 1,3 \right \} or \left \{ 2,4 \right \}$, use the result with basis  $\left \{X,X,Y,Y\right \}$, $\left \{X,X,Z,Z\right \}$, $\left \{Y,Y,X,X\right \}$, $\left \{Y,Y,Z,Z\right \}$, $\left \{Z,Z,X,X\right \}$, $\left \{Z,Z,Y,Y\right \}$. For example, 
\begin{widetext}
\begin{align}
S_{1,2}^{\left \{ 2,4 \right \} } = S_{0,1,0,2}
& = p_{1}^{\left\{ X,X,Y,Y \right \}} - p_{2}^{\left\{ X,X,Y,Y \right \}} + p_{3}^{\left\{ X,X,Y,Y \right \}} - p_{4}^{\left\{ X,X,Y,Y \right \}}\notag\\
& - p_{5}^{\left\{ X,X,Y,Y \right \}} + p_{6}^{\left\{ X,X,Y,Y \right \}} - p_{7}^{\left\{ X,X,Y,Y \right \}} + p_{8}^{\left\{ X,X,Y,Y \right \}}\notag\\
& + p_{9}^{\left\{ X,X,Y,Y \right \}} - p_{10}^{\left\{ X,X,Y,Y \right \}} + p_{11}^{\left\{ X,X,Y,Y \right \}} - p_{12}^{\left\{ X,X,Y,Y \right \}}\notag\\
& - p_{13}^{\left\{ X,X,Y,Y \right \}} + p_{14}^{\left\{ X,X,Y,Y \right \}} - p_{15}^{\left\{ X,X,Y,Y \right \}} + p_{16}^{\left\{ X,X,Y,Y \right \}}
\end{align}
\end{widetext}

\item[c.]
Similarly, For $S_{i_1,i_2}^{\left \{ x_1,x_2 \right \} }$ with $i_1 \ne i_2$, $\left \{ x_1,x_2 \right \} = \left \{ 1,2 \right \} or \left \{ 3,4 \right \}$, use the result with basis  $\left \{X,Y,X,Y\right \}$, $\left \{X,Z,X,Z\right \}$, $\left \{Y,X,Y,X\right \}$, $\left \{Y,Z,Y,Z\right \}$, $\left \{Z,X,Z,X\right \}$, $\left \{Z,Y,Z,Y\right \}$. 

\item[d.]
For $S_{i_1,i_2}^{\left \{ x_1,x_2 \right \} }$ with $i_1 \ne i_2$, $\left \{ x_1,x_2 \right \} = \left \{ 1,4 \right \} or \left \{ 2,3 \right \}$, there are two calculation results corresponding to these values. In principle they give same values, here we take the average value. For example,

\begin{equation}
S_{1,2}^{\left \{ 1,4 \right \} } = \frac{S_{1,0,0,2}^{\left \{X,Y,X,Y\right \}} + S_{1,0,0,2}^{\left \{ X,X,Y,Y\right \}}}{2}
\end{equation}

\item[e.]
For all one-fold terms, where $S_{i_1,i_2}^{\left \{ x_1,x_2 \right \} }$ with $i_1 or i_2 =0$, there will be 5 corresponding calculation results. We take the average value of all of them. For example,
\begin{widetext}
\begin{equation}
S_{1,0}^{\left \{ 1,2 \right \} } = S_{1,0}^{\left \{ 1,3 \right \} } = S_{1,0}^{\left \{ 1,4 \right \} } = \frac{S_{1,0,0,0}^{\left \{X,Y,X,Y\right \}} + S_{1,0,0,0}^{\left \{ X,X,Y,Y\right \}} + S_{1,0,0,0}^{\left \{ X,X,Z,Z \right \}} +S_{1,0,0,0}^{\left \{ X,Z,X,Z\right \}} + S_{1,0,0,0}^{\left \{ X,X,X,X \right \}}}{5}
\end{equation}
\end{widetext}

\item[f.]
$S_{0,0}^{\left \{ x_1,x_2 \right \} } = 1$ for any $\left \{ x_1,x_2 \right \} $
\end{itemize}

So far we've calculated all $S_{i_1,i_2}^{\left \{ x_1,x_2 \right \} }$ in Eq(20), density matrices of all six 2-qubit subsystems can be reconstructed.

\newpage

\section{Measurement Results}

\subsection{State Fidelities of Reconstructed Density Matrices of State ${\psi'}$}

\begin{figure}[H]
\centering
\includegraphics[width=0.9\linewidth]{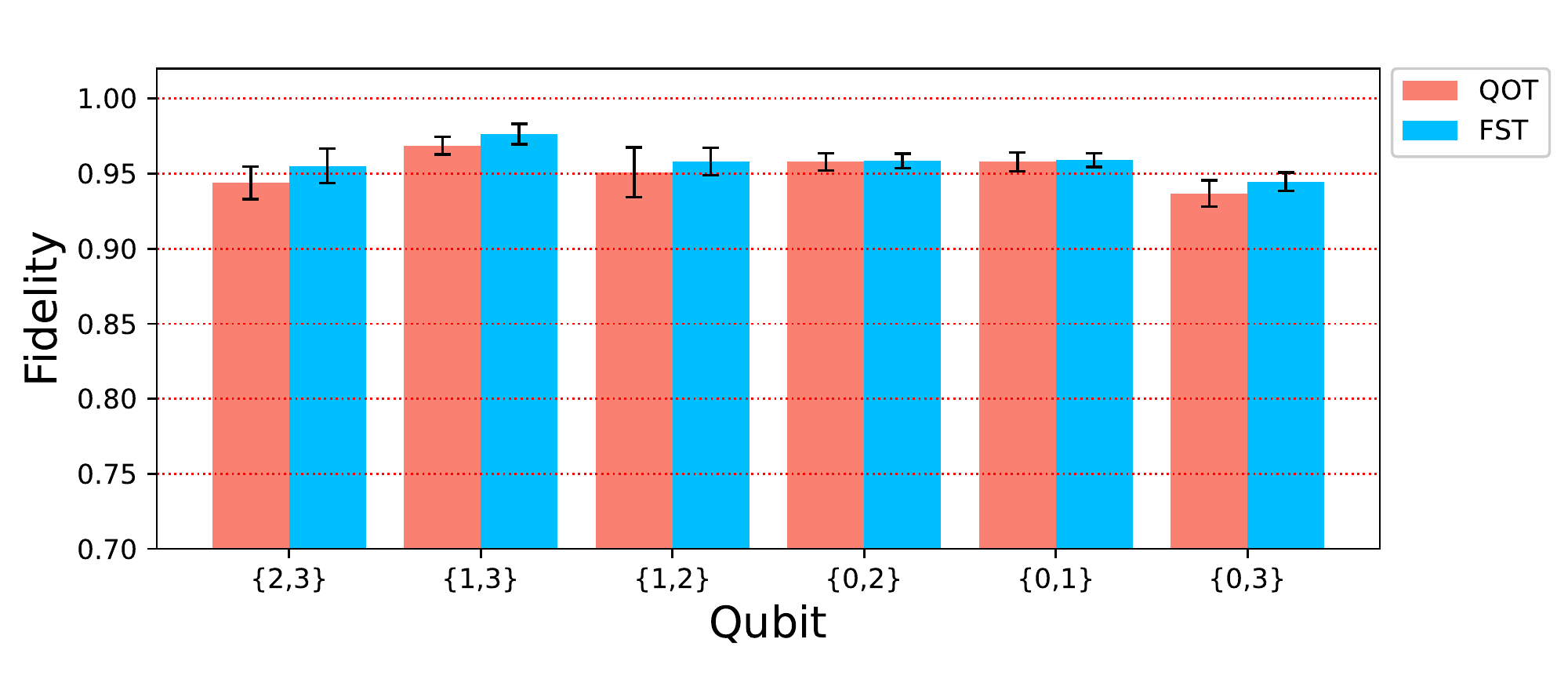}
\caption{2-qubit state fidelity { with ${\psi'}_{i_1,i_2}^R$ of} ${\psi'}_{i_1,i_2}^F$ and ${\psi'}_{i_1,i_2}^O$, {error bars show 95\% confidence interval}}
\end{figure}


\subsection{Fidelities between Density Matrices Reconstructed by QOT and FST}

\begin{figure}[H]
\centering
\includegraphics[width=0.9\linewidth]{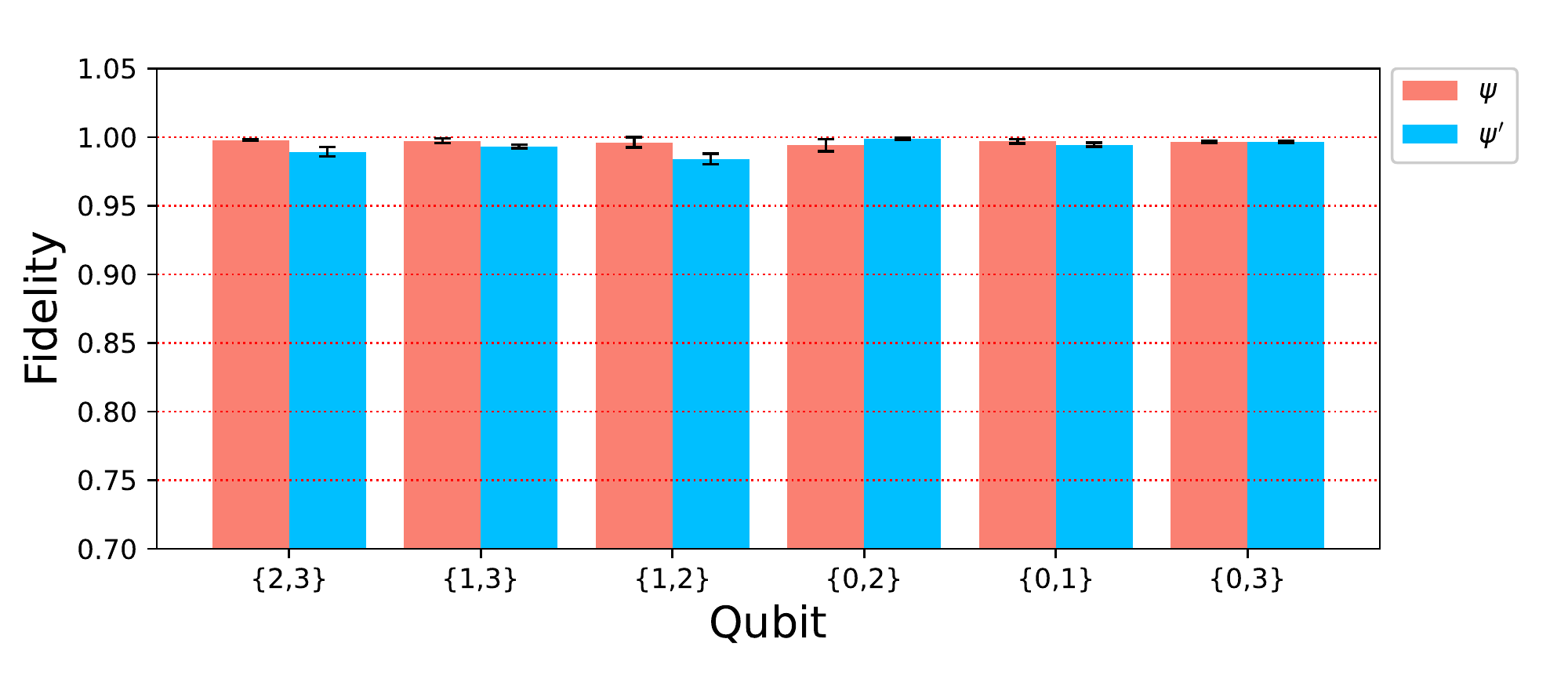}
\caption{Fidelities between density matrices reconstructed by QOT and FST. Red bars show results of state ${\psi}$. Blue bars show results of state ${\psi}'$. Error bars show 95\% confidence intervals.}
\end{figure}

\newpage

\subsection{Estimation of Density Matrix of State ${\psi}$}

\begin{figure}[H]
\centering
\includegraphics[width=0.65\linewidth]{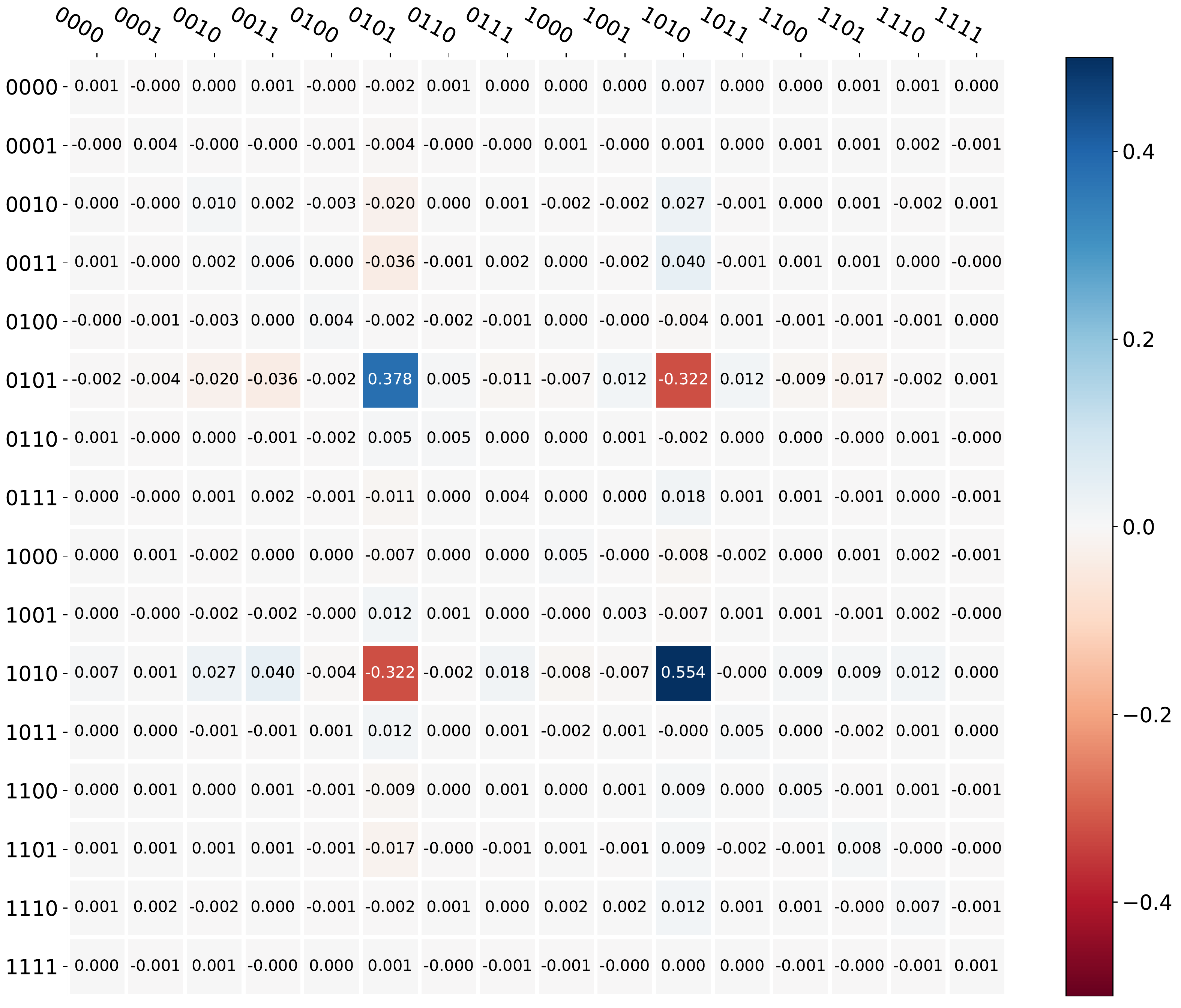}
\caption{Real part of density matrix of ${\psi}^F$  obtained by FST.}
\end{figure}

\begin{figure}[H]
\centering
\includegraphics[width=0.65\linewidth]{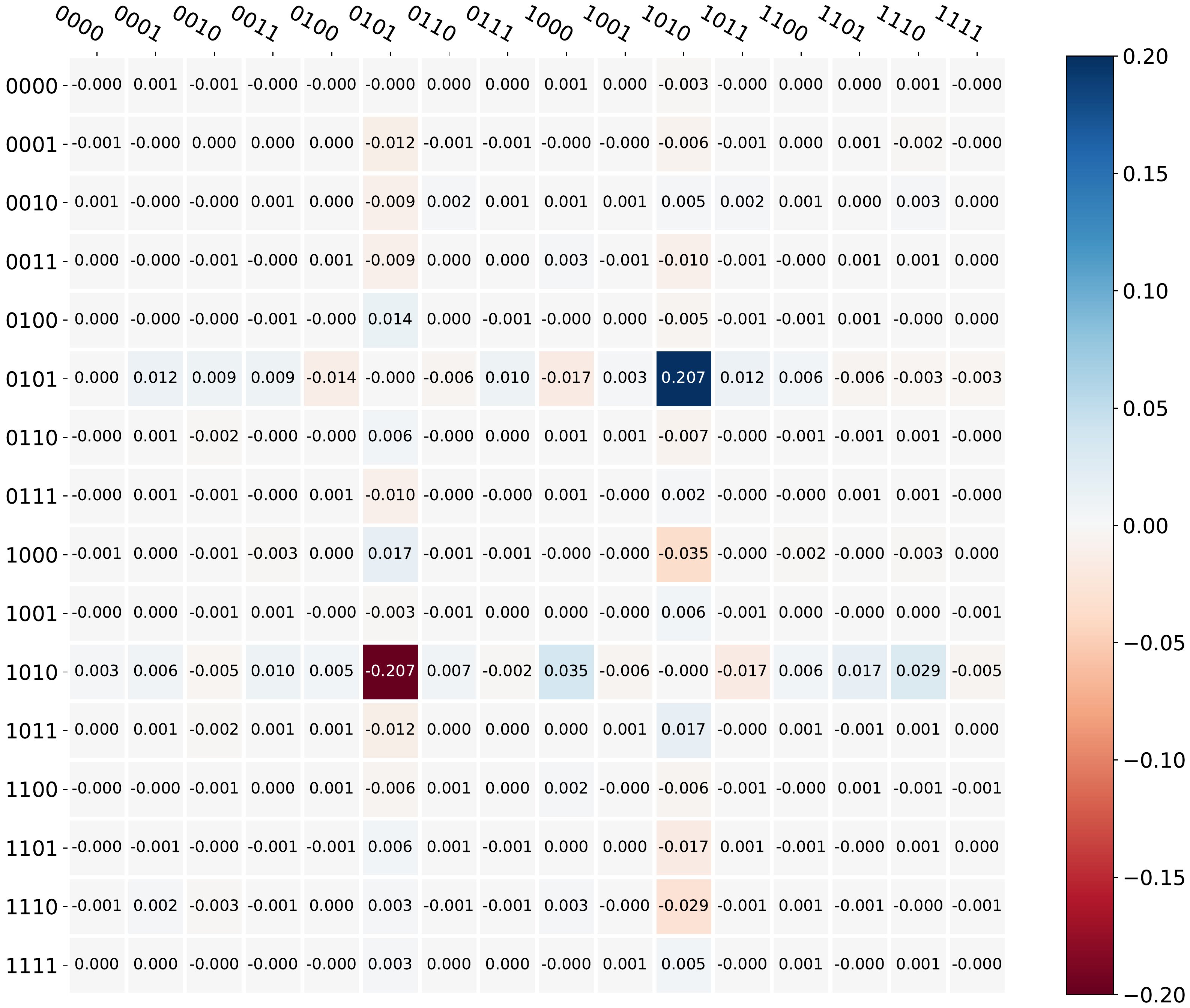}
\caption{Imaginary part of density matrix of ${\psi}^F$  obtained by FST.}
\end{figure}


\subsection{Estimation of Density Matrices of Subsystems of ${\psi}$}

\begin{figure}[H]
\centering
\includegraphics[width=\linewidth]{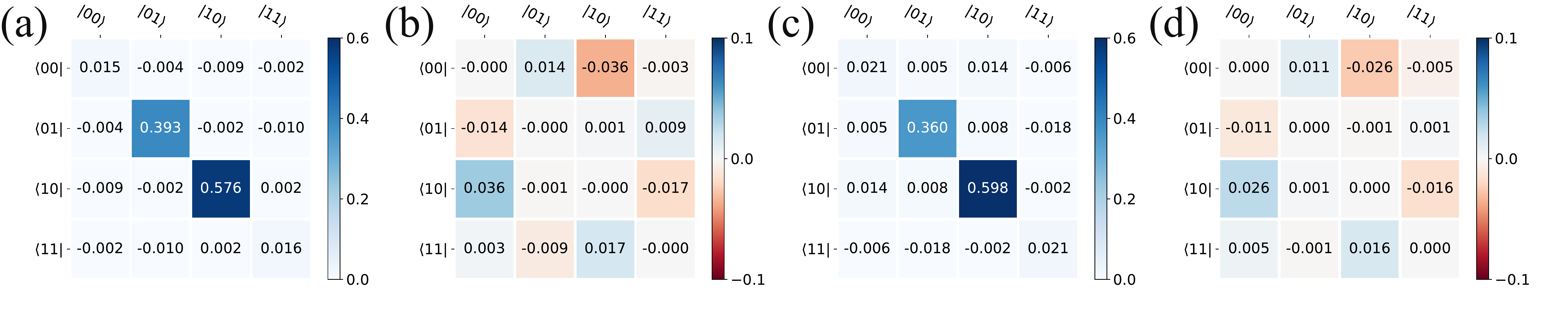}
\caption{(a) real and (b) imaginary part of density matrix of 2-qubit subsystem ${\psi}_{2,3}^F$  obtained by FST (c) real and (d) imaginary part of density matrix of 2-qubit subsystem ${\psi}_{2,3}^O$ obtained by QOT.}
\end{figure}

\begin{figure}[H]
\centering
\includegraphics[width=\linewidth]{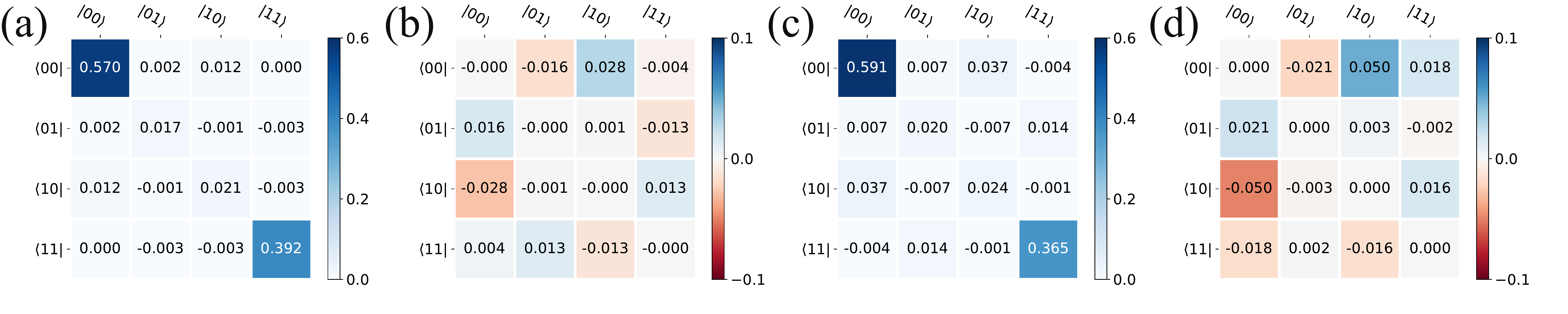}
\caption{(a) real and (b) imaginary part of density matrix of 2-qubit subsystem ${\psi}_{1,3}^F$  obtained by FST (c) real and (d) imaginary part of density matrix of 2-qubit subsystem ${\psi}_{1,3}^O$ obtained by QOT.}
\end{figure}

\begin{figure}[H]
\centering
\includegraphics[width=\linewidth]{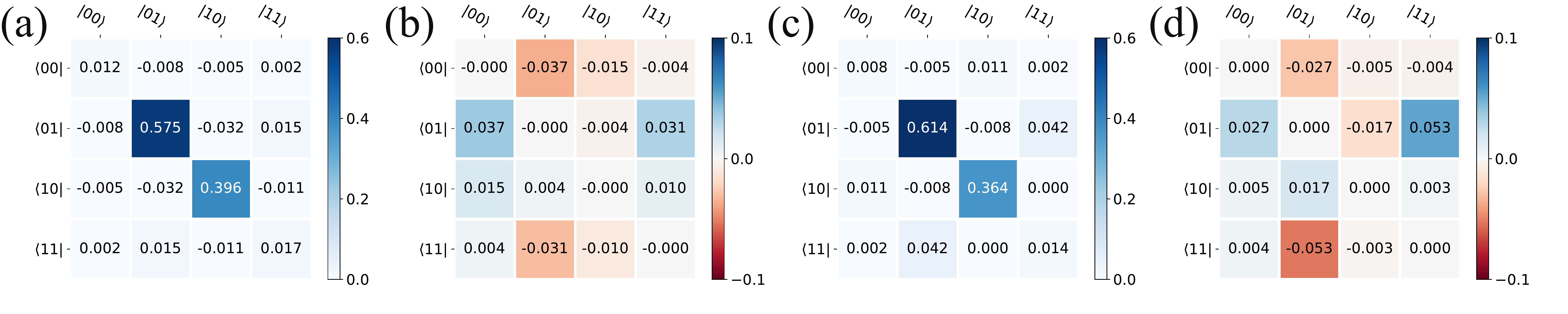}
\caption{(a) real and (b) imaginary part of density matrix of 2-qubit subsystem ${\psi}_{1,2}^F$  obtained by FST(c) real and (d) imaginary part of density matrix of 2-qubit subsystem ${\psi}_{1,2}^O$ obtained by QOT.}
\end{figure}

\begin{figure}[H]
\centering
\includegraphics[width=\linewidth]{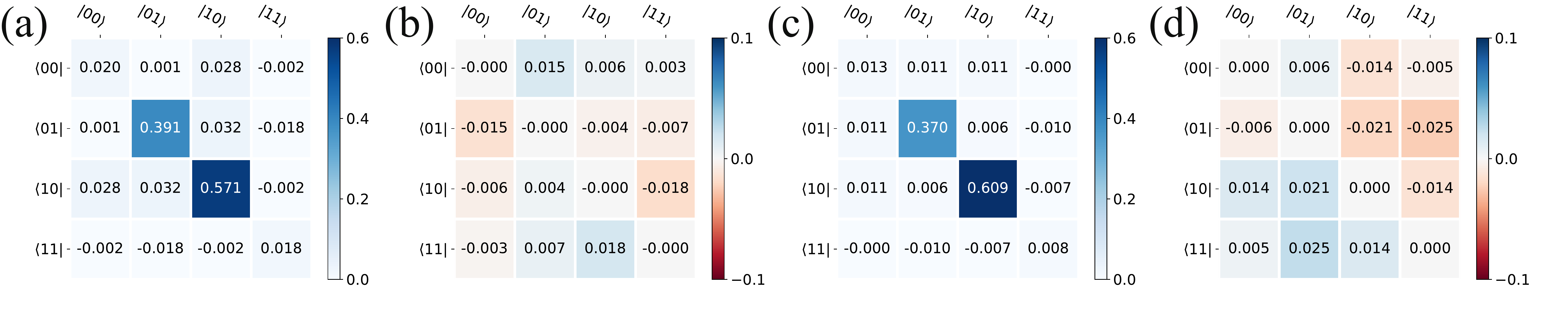}
\caption{(a) real and (b) imaginary part of density matrix of 2-qubit subsystem ${\psi}_{0,3}^F$  obtained by FST (c) real and (d) imaginary part of density matrix of 2-qubit subsystem ${\psi}_{0,3}^O$ obtained by QOT.}
\end{figure}

\begin{figure}[H]
\centering
\includegraphics[width=\linewidth]{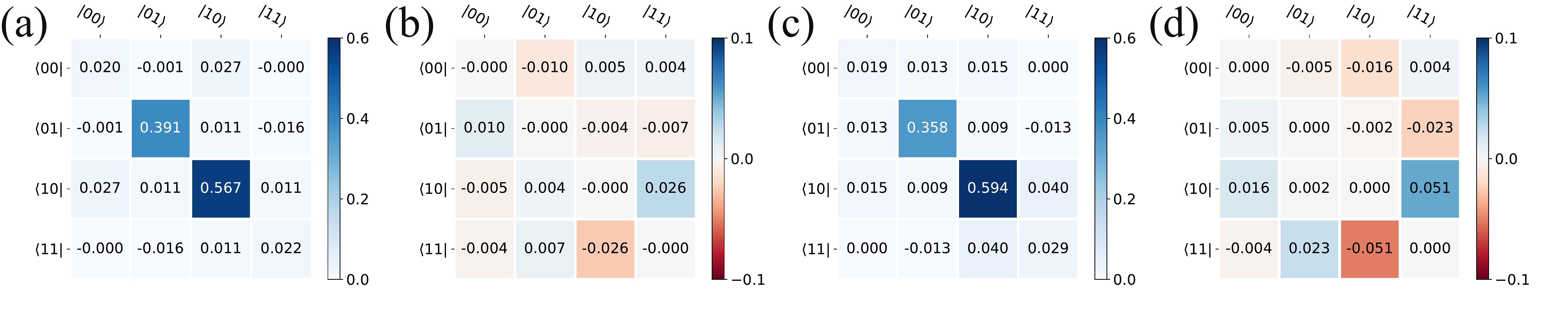}
\caption{(a) real and (b) imaginary part of density matrix of 2-qubit subsystem ${\psi}_{0,1}^F$  obtained by FST (c) real and (d) imaginary part of density matrix of 2-qubit subsystem ${\psi}_{0,1}^O$ obtained by QOT.}
\end{figure}

\begin{figure}[H]
\centering
\includegraphics[width=\linewidth]{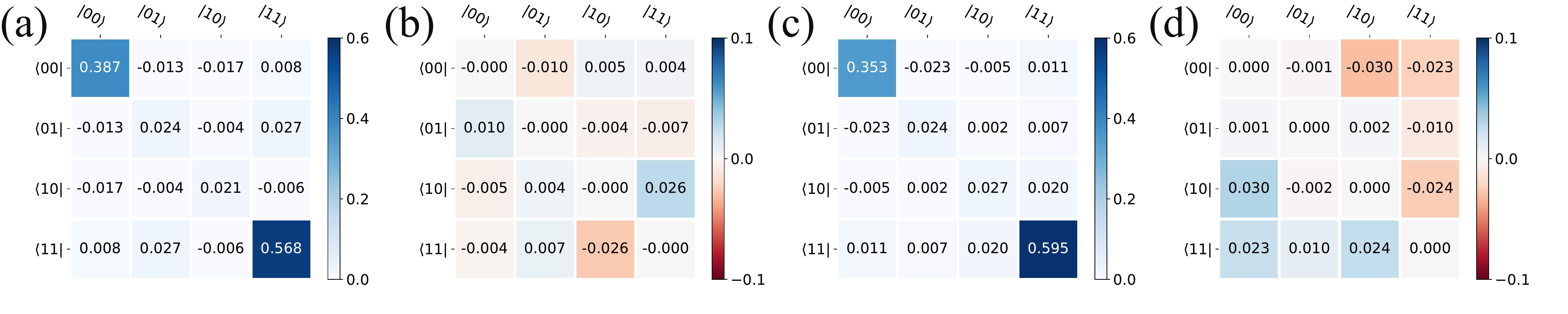}
\caption{(a) real and (b) imaginary part of density matrix of 2-qubit subsystem ${\psi}_{0,2}^F$  obtained by FST (c) real and (d) imaginary part of density matrix of 2-qubit subsystem ${\psi}_{0,2}^O$ obtained by QOT.}
\end{figure}

\subsection{Estimation of Density Matrix of State ${\psi'}$}
\begin{figure}[H]
\centering
\includegraphics[width=0.65\linewidth]{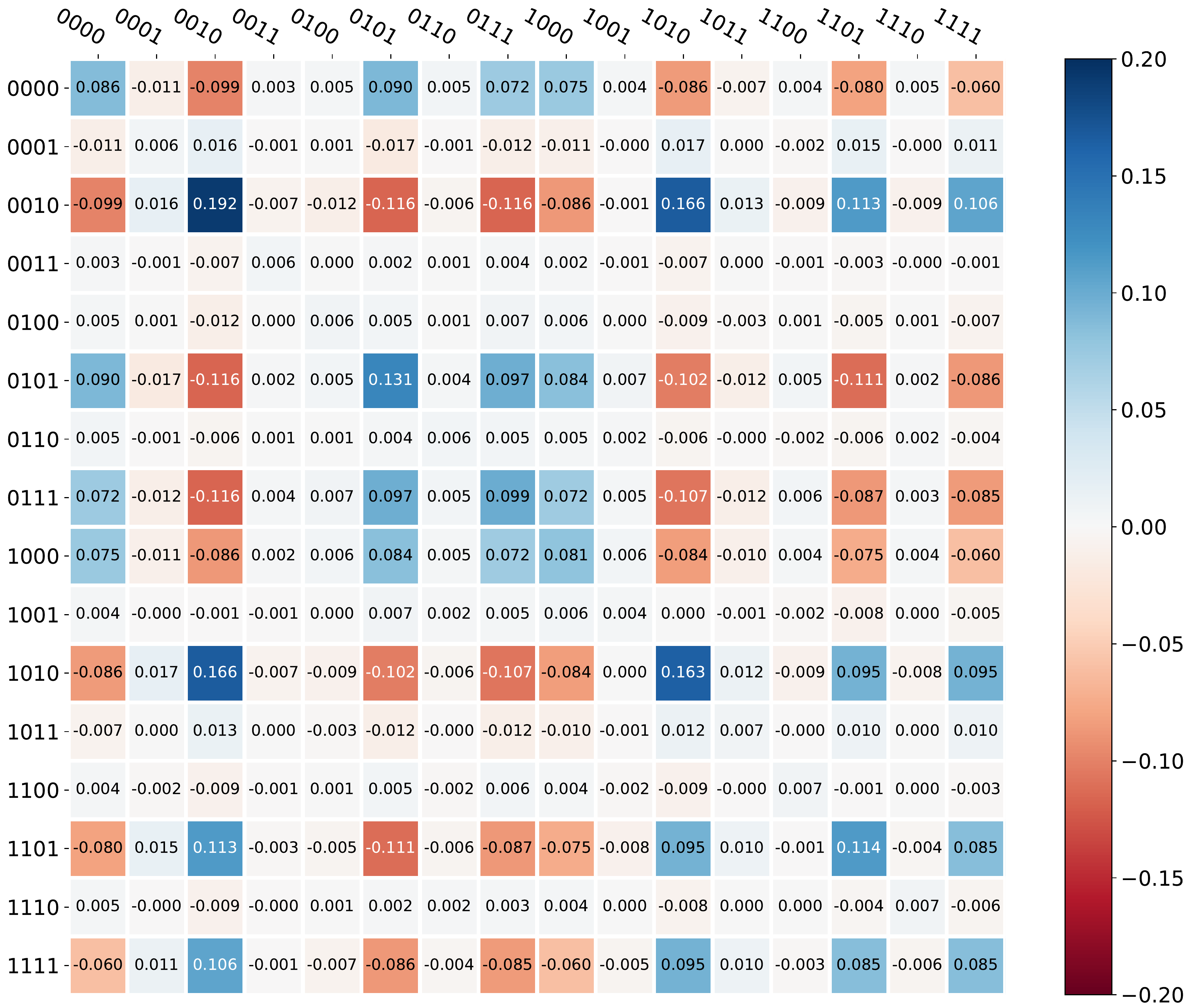}
\caption{Real part of density matrix of ${\psi'}^F$  obtained by FST.}
\end{figure}

\begin{figure}[H]
\centering
\includegraphics[width=0.65\linewidth]{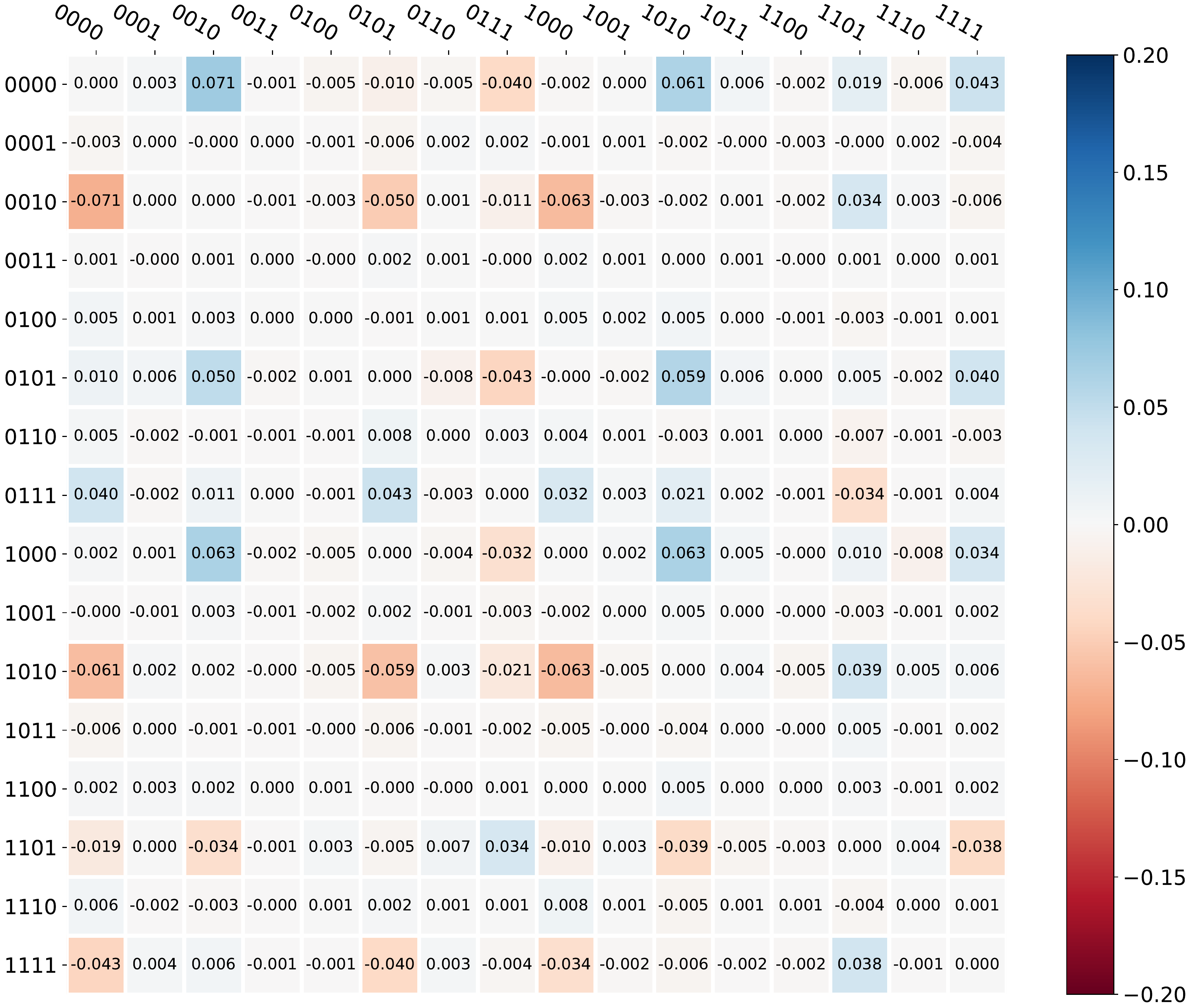}
\caption{Imaginary part of density matrix of ${\psi'}^F$  obtained by FST.}
\end{figure}


\subsection{Estimation of Density Matrices of Subsystems of ${\psi'}$}

\begin{figure}[H]
\centering
\includegraphics[width=\linewidth]{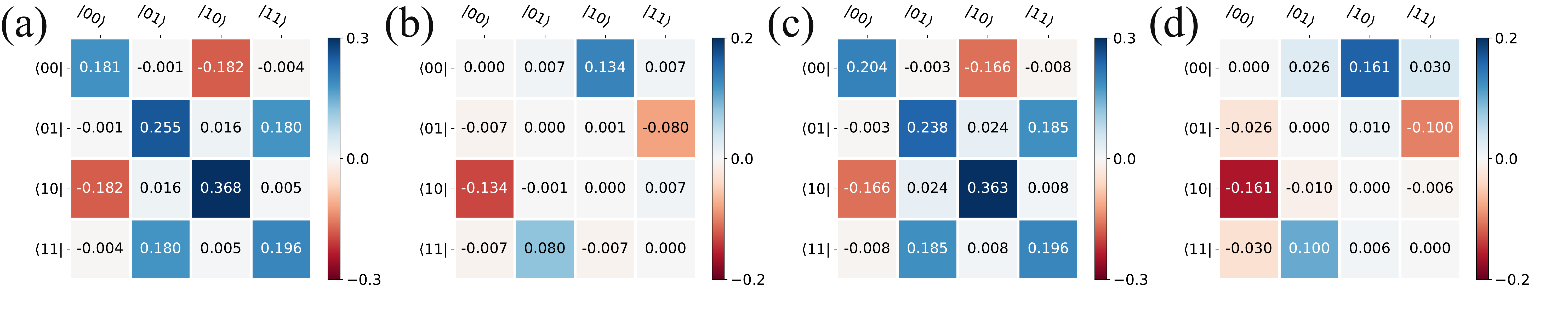}
\caption{(a) real and (b) imaginary part of density matrix of 2-qubit subsystem ${\psi'}_{2,3}^F$  obtained by FST (c) real and (d) imaginary part of density matrix of 2-qubit subsystem ${\psi'}_{2,3}^O$ obtained by QOT.}
\end{figure}

\begin{figure}[H]
\centering
\includegraphics[width=\linewidth]{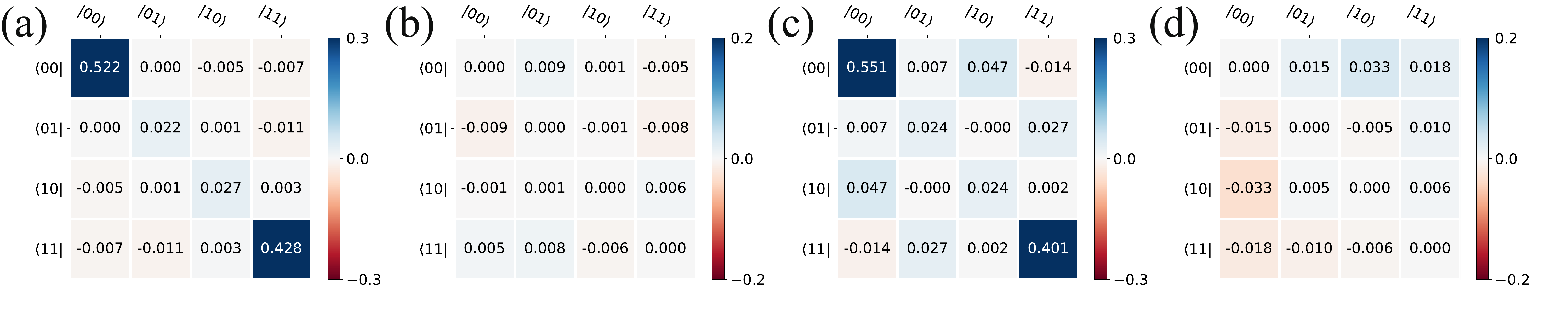}
\caption{(a) real and (b) imaginary part of density matrix of 2-qubit subsystem ${\psi'}_{1,3}^F$  obtained by FST (c) real and (d) imaginary part of density matrix of 2-qubit subsystem ${\psi'}_{1,3}^O$ obtained by QOT.}
\end{figure}

\begin{figure}[H]
\centering
\includegraphics[width=\linewidth]{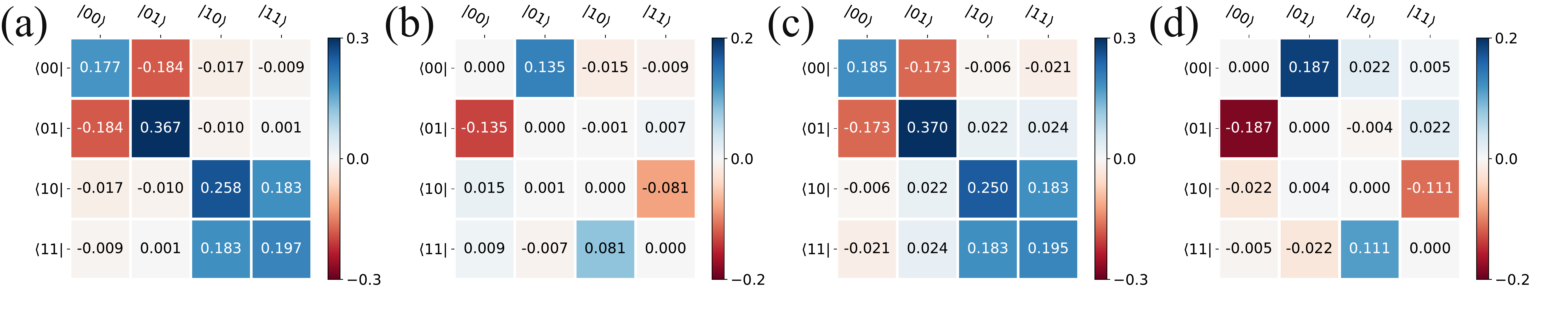}
\caption{(a) real and (b) imaginary part of density matrix of 2-qubit subsystem ${\psi'}_{1,2}^F$  obtained by FST(c) real and (d) imaginary part of density matrix of 2-qubit subsystem ${\psi'}_{1,2}^O$ obtained by QOT.}
\end{figure}

\begin{figure}[H]
\centering
\includegraphics[width=\linewidth]{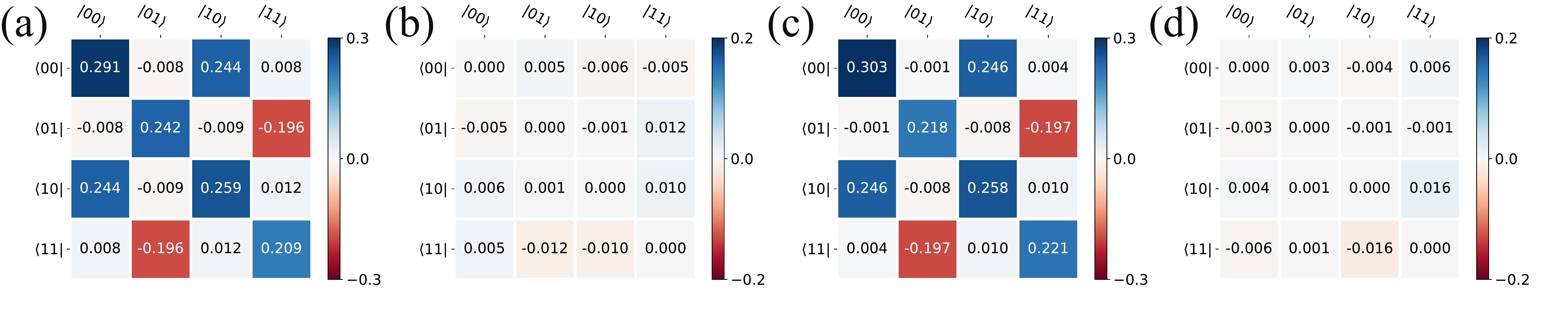}
\caption{(a) real and (b) imaginary part of density matrix of 2-qubit subsystem ${\psi'}_{0,3}^F$  obtained by FST (c) real and (d) imaginary part of density matrix of 2-qubit subsystem ${\psi'}_{0,3}^O$ obtained by QOT.}
\end{figure}

\begin{figure}[H]
\centering
\includegraphics[width=\linewidth]{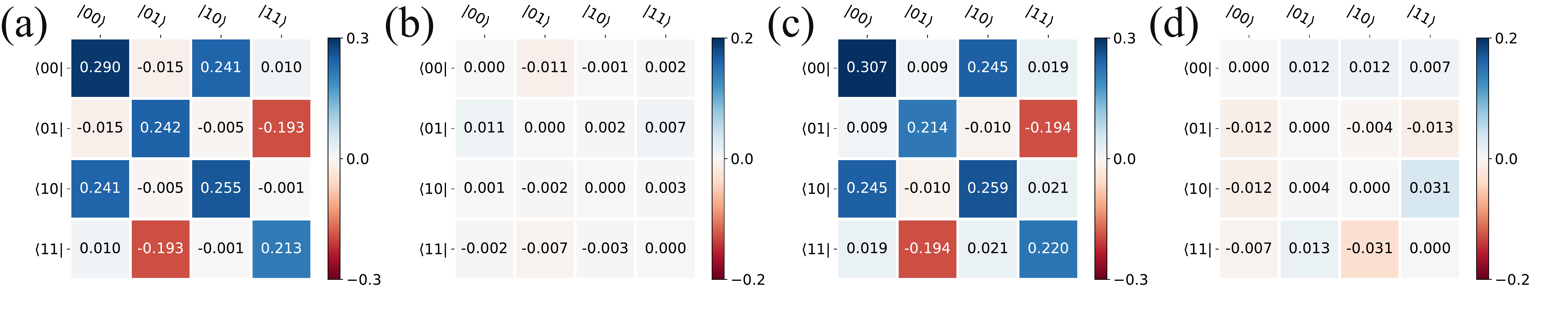}
\caption{(a) real and (b) imaginary part of density matrix of 2-qubit subsystem ${\psi'}_{0,1}^F$  obtained by FST (c) real and (d) imaginary part of density matrix of 2-qubit subsystem ${\psi'}_{0,1}^O$ obtained by QOT.}
\end{figure}

\begin{figure}[H]
\centering
\includegraphics[width=\linewidth]{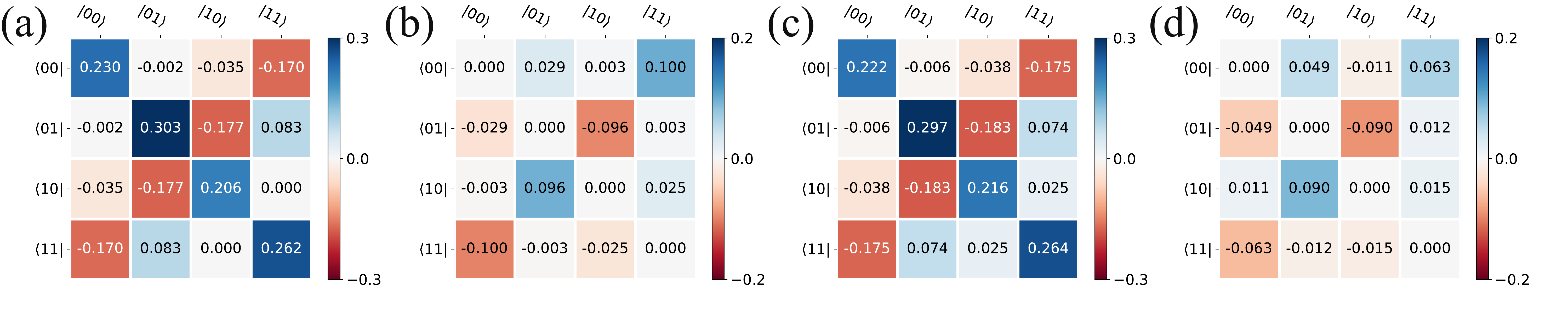}
\caption{(a) real and (b) imaginary part of density matrix of 2-qubit subsystem ${\psi'}_{0,2}^F$  obtained by FST (c) real and (d) imaginary part of density matrix of 2-qubit subsystem ${\psi'}_{0,2}^O$ obtained by QOT.}
\end{figure}

\newpage

\section{six-qubit QOT experiment}

\subsection{Overview}

\begin{figure}[H]
\centering
\includegraphics[width=0.8\linewidth]{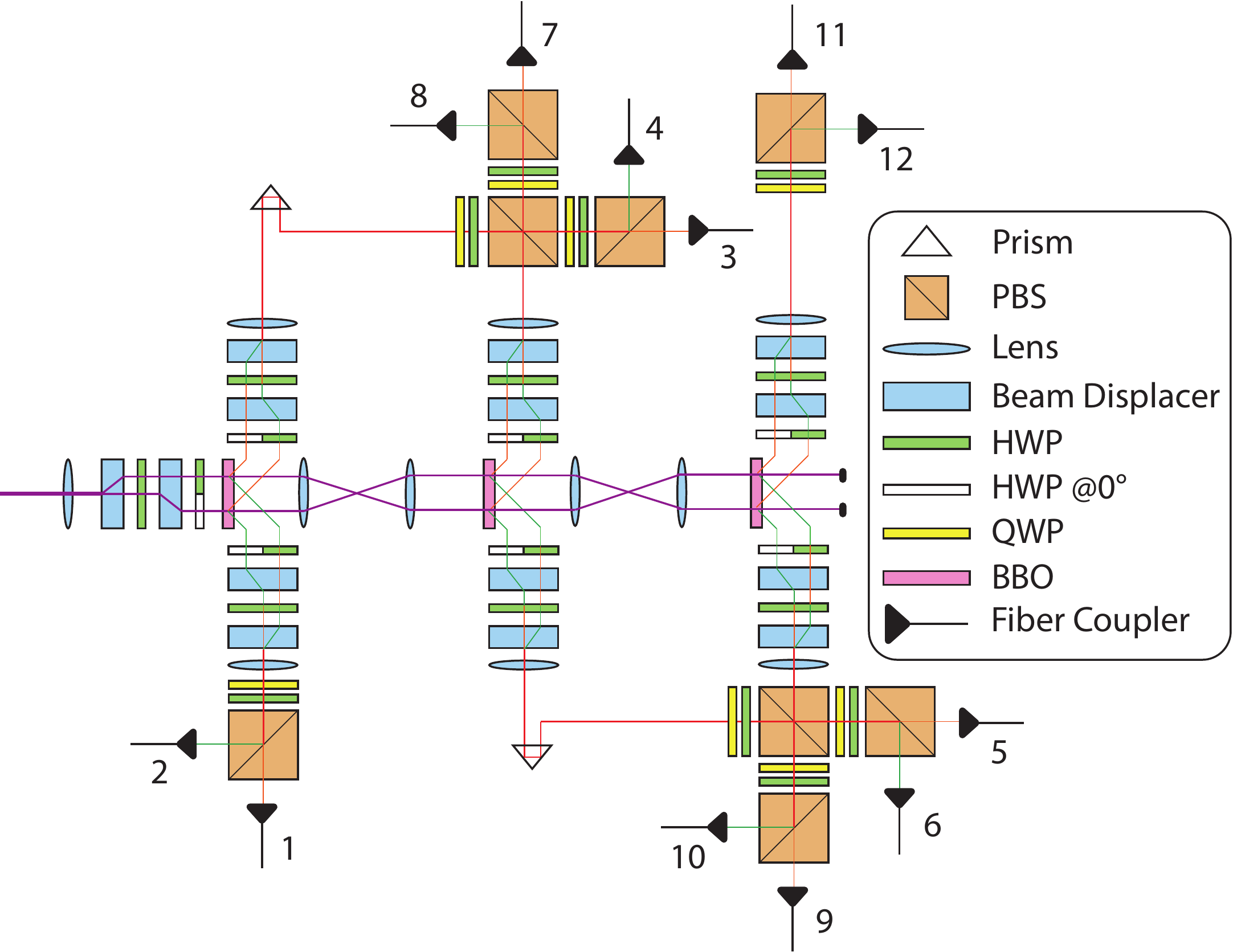}
\caption{Schematic of experimental set-up for generating 6-photon entanglement, with the detectors labeled by order.\label{fig:6-set}}
\end{figure}

Following the scheme discussed in Sec. III, we scaled up the system to 6 photons. Fig.\ref{fig:6-set} shows the experimental set-up for 6-photon entanglement. In this case, we attempt to genearate the state ${\psi}_{{GHZ}_6}^R= \frac{1}{\sqrt{2}} (\ket{HVHVHV} + \ket{VHVHVH})$ and perform QOT on it. In the experiment, the 6-photon events are recorded at an average count rate 0.05 Hz. The QOT measurement takes around 80 hours to record all $700 \times 21 = 14700$ 6-photon events. Subsection B show the results of the 6-qubit QOT experiment. 

We can notice that the state fidelities shown in the main text are considerably lower than the fidelities obtained in 4-qubit QOT. In our view, there are 3 major factors contribute to this situation:

Firstly, since FST is not feasible in this case, it is hard to confirm the exact state to be compared with the results. Even though we set the system to generate a reference state  ${\psi}_{{GHZ}_6}^R$, some optical elements can introduce systematic errors which is hard to predict before experiment. 

Besides, as we try to improve the multi-photon event count rate to ensure the 6-photon measurement feasible, we used a broader bandpass to filter the signal photons, which would allow more frequency correlation and reduce the polarization correlation. As a result, the fidelity of the 2-photon entangled state generated in this case would be lower than the fidelity in the 4-photon experiment. This difference makes a major contribution to the low state fidelity.

Finally, due to the low multi-photon event count rate in this case, the system inconsistency and the random coincidence caused by noise became more significant in the final result. The low count rate also forced us to take fewer measurements in limited time as well, which leads to a worse statistical significance.

By applying QOT to such a experimental system, we can diagnose and understand those systematic errors within acceptable time. We think it will be another useful application of QOT in quantum optics research.

\newpage

\subsection{Reconstructed 2-qubit Subsystem Density Matrices}

\twocolumngrid

\begin{figure}[H]
\centering
\includegraphics[width=\linewidth]{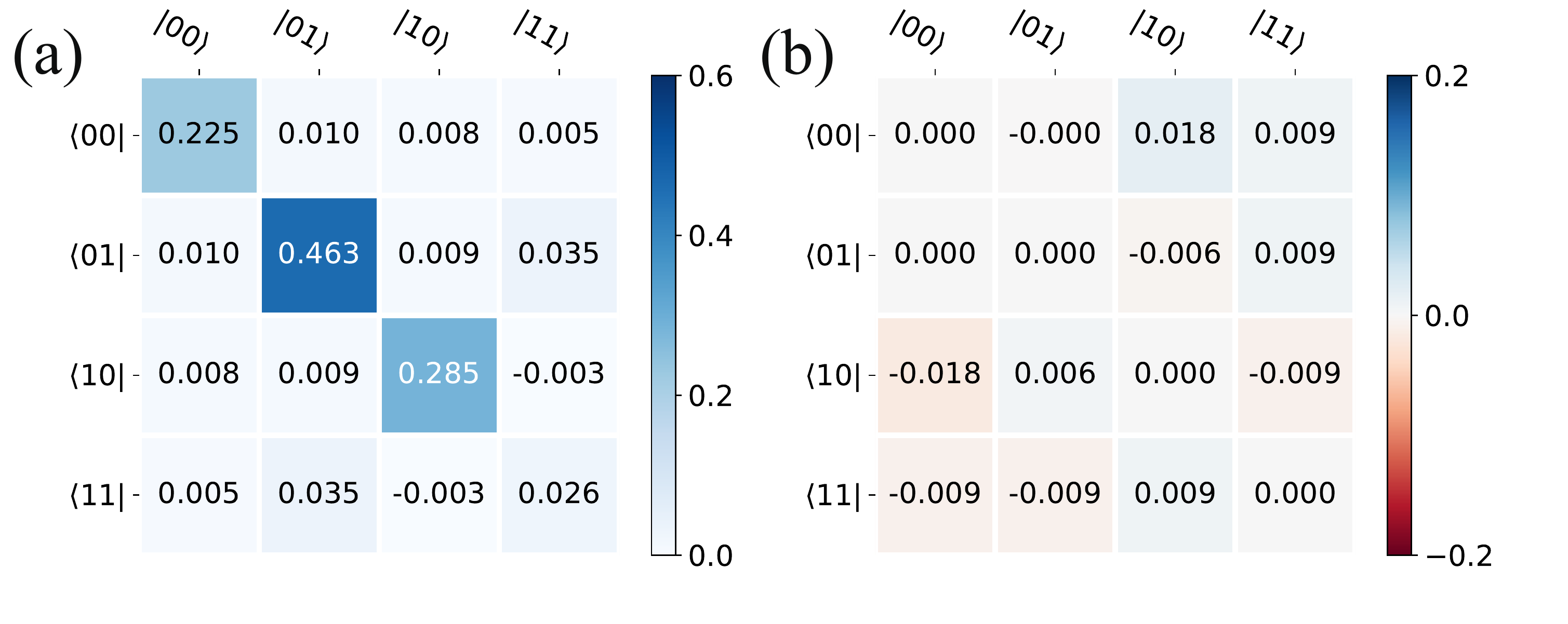}
\caption{(a) real and (b) imaginary part of density matrix of 2-qubit subsystem ${\psi^{(6)}}_{0,1}$  obtained by QOT}
\end{figure}
\begin{figure}[H]
\centering
\includegraphics[width=\linewidth]{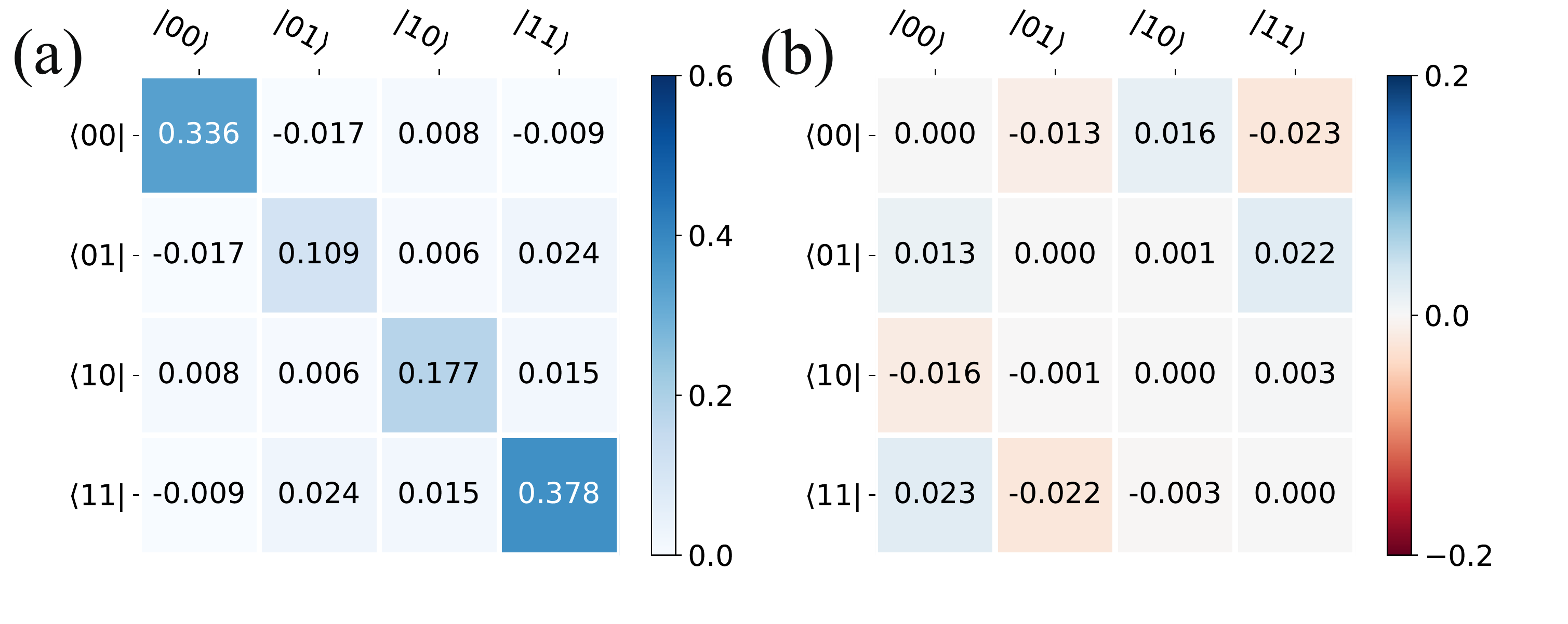}
\caption{(a) real and (b) imaginary part of density matrix of 2-qubit subsystem ${\psi^{(6)}}_{0,2}$  obtained by QOT}
\end{figure}
\begin{figure}[H]
\centering
\includegraphics[width=\linewidth]{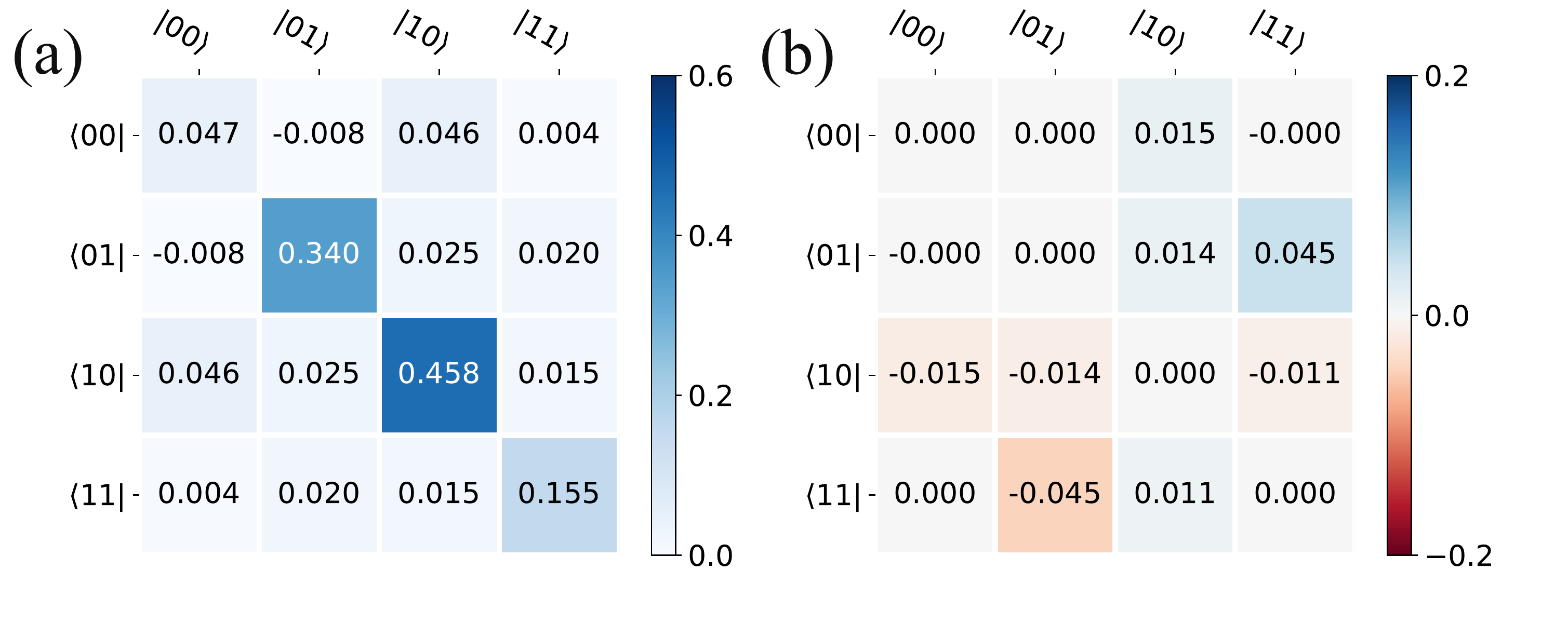}
\caption{(a) real and (b) imaginary part of density matrix of 2-qubit subsystem ${\psi^{(6)}}_{0,3}$  obtained by QOT}
\end{figure}
\begin{figure}[H]
\centering
\includegraphics[width=\linewidth]{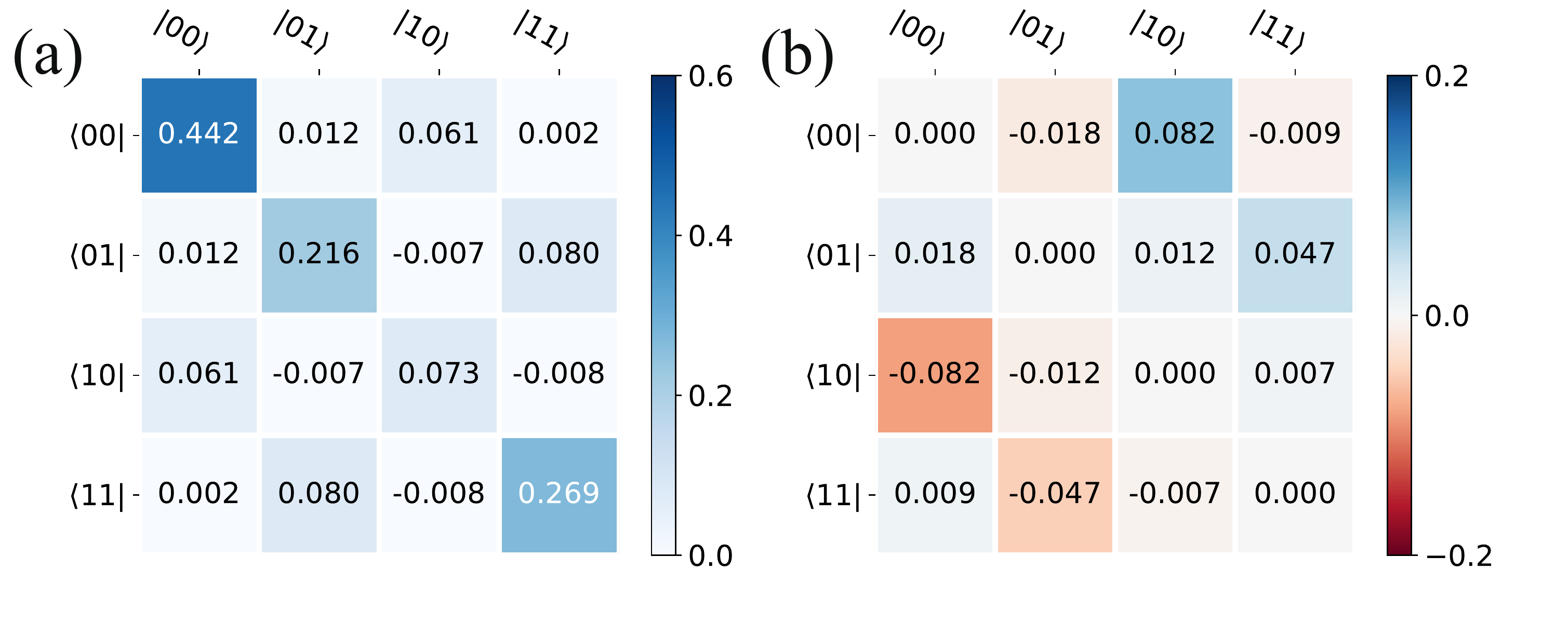}
\caption{(a) real and (b) imaginary part of density matrix of 2-qubit subsystem ${\psi^{(6)}}_{0,4}$  obtained by QOT}
\end{figure}
\begin{figure}[H]
\centering
\includegraphics[width=\linewidth]{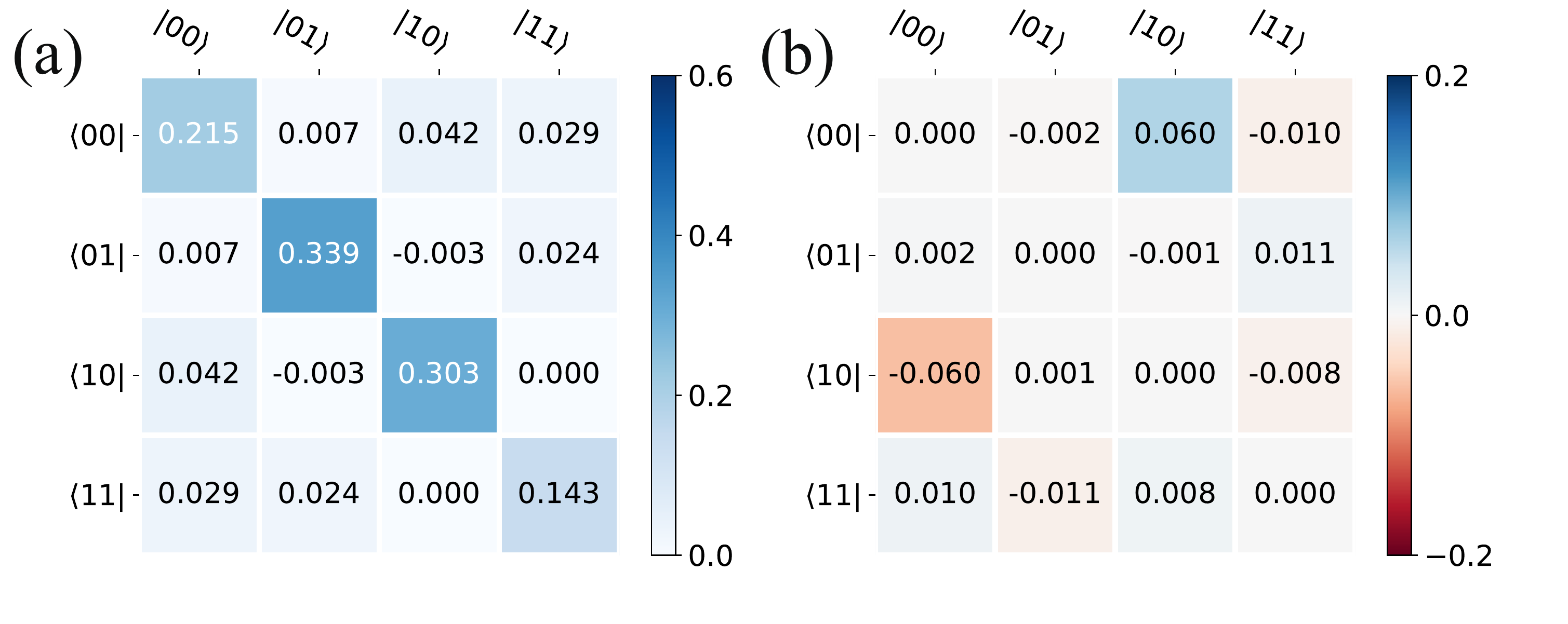}
\caption{(a) real and (b) imaginary part of density matrix of 2-qubit subsystem ${\psi^{(6)}}_{0,5}$  obtained by QOT}
\end{figure}
\begin{figure}[H]
\centering
\includegraphics[width=\linewidth]{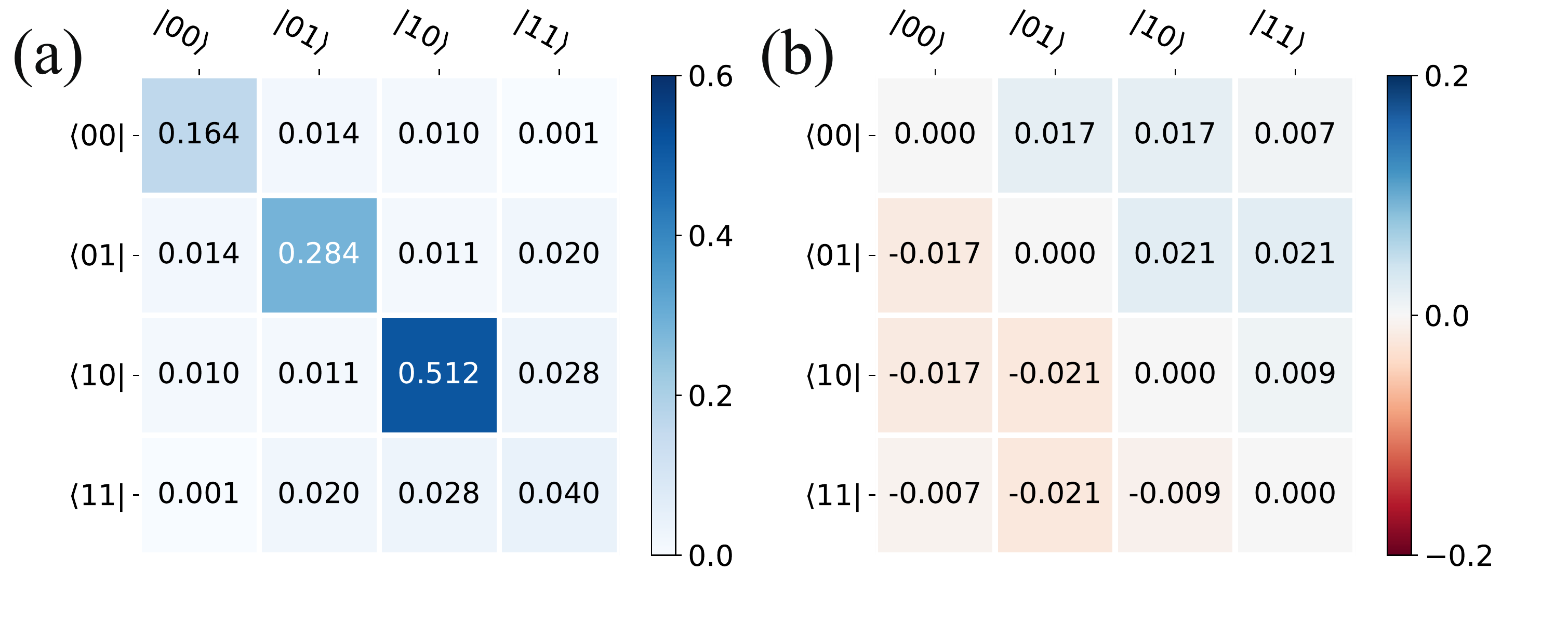}
\caption{(a) real and (b) imaginary part of density matrix of 2-qubit subsystem ${\psi^{(6)}}_{1,2}$  obtained by QOT}
\end{figure}
\begin{figure}[H]
\centering
\includegraphics[width=\linewidth]{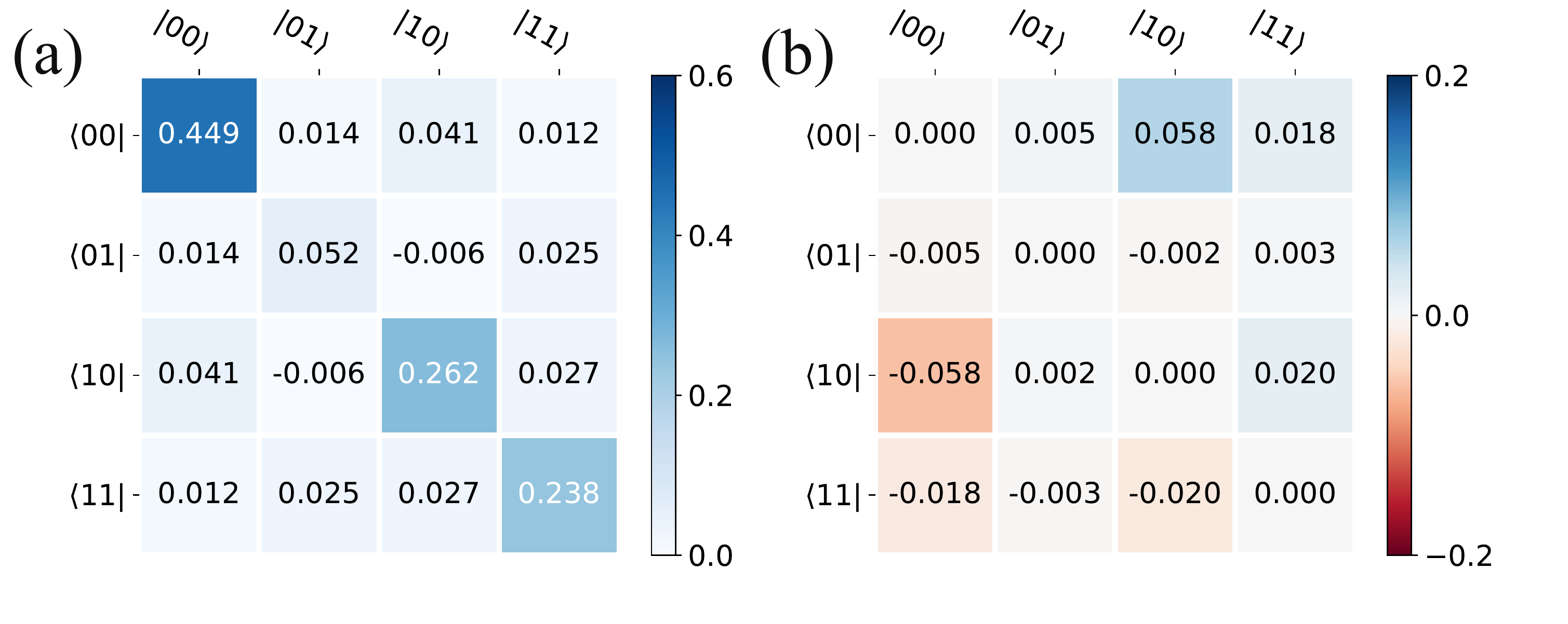}
\caption{(a) real and (b) imaginary part of density matrix of 2-qubit subsystem ${\psi^{(6)}}_{1,3}$  obtained by QOT}
\end{figure}
\begin{figure}[H]
\centering
\includegraphics[width=\linewidth]{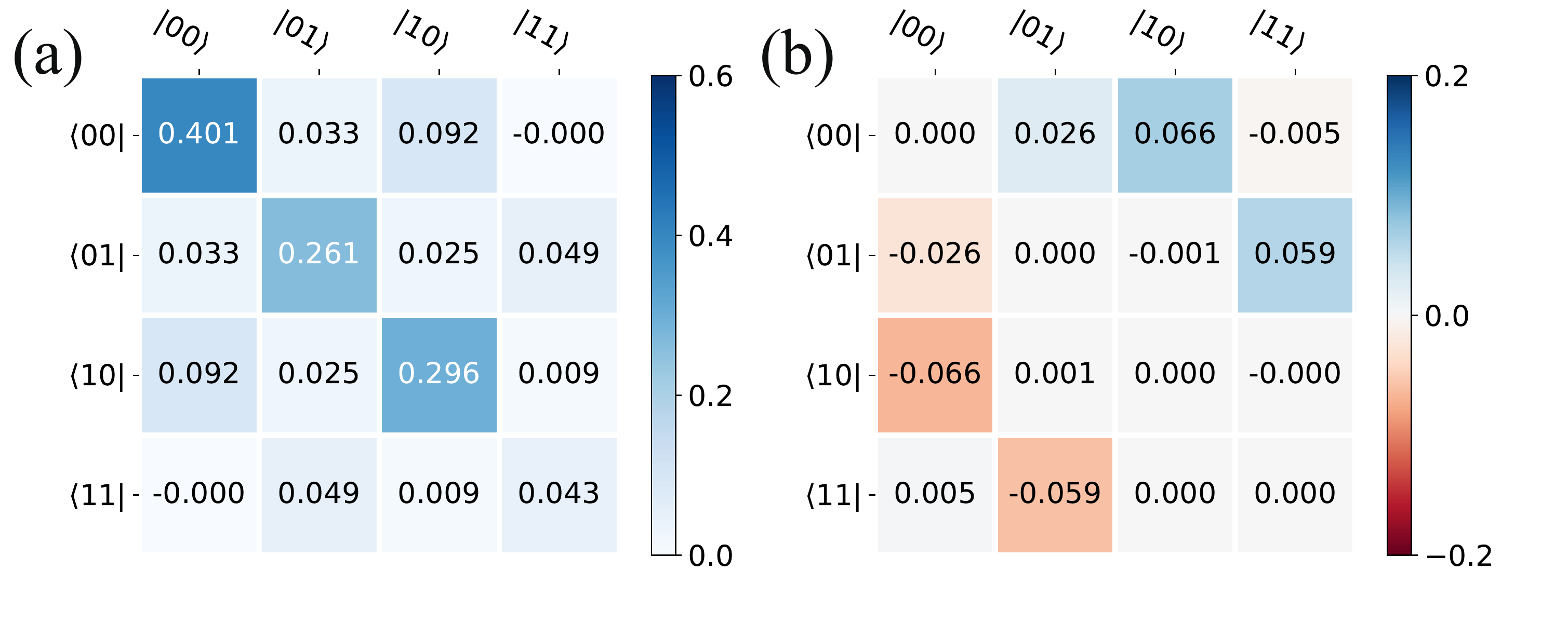}
\caption{(a) real and (b) imaginary part of density matrix of 2-qubit subsystem ${\psi^{(6)}}_{1,4}$  obtained by QOT}
\end{figure}
\begin{figure}[H]
\centering
\includegraphics[width=\linewidth]{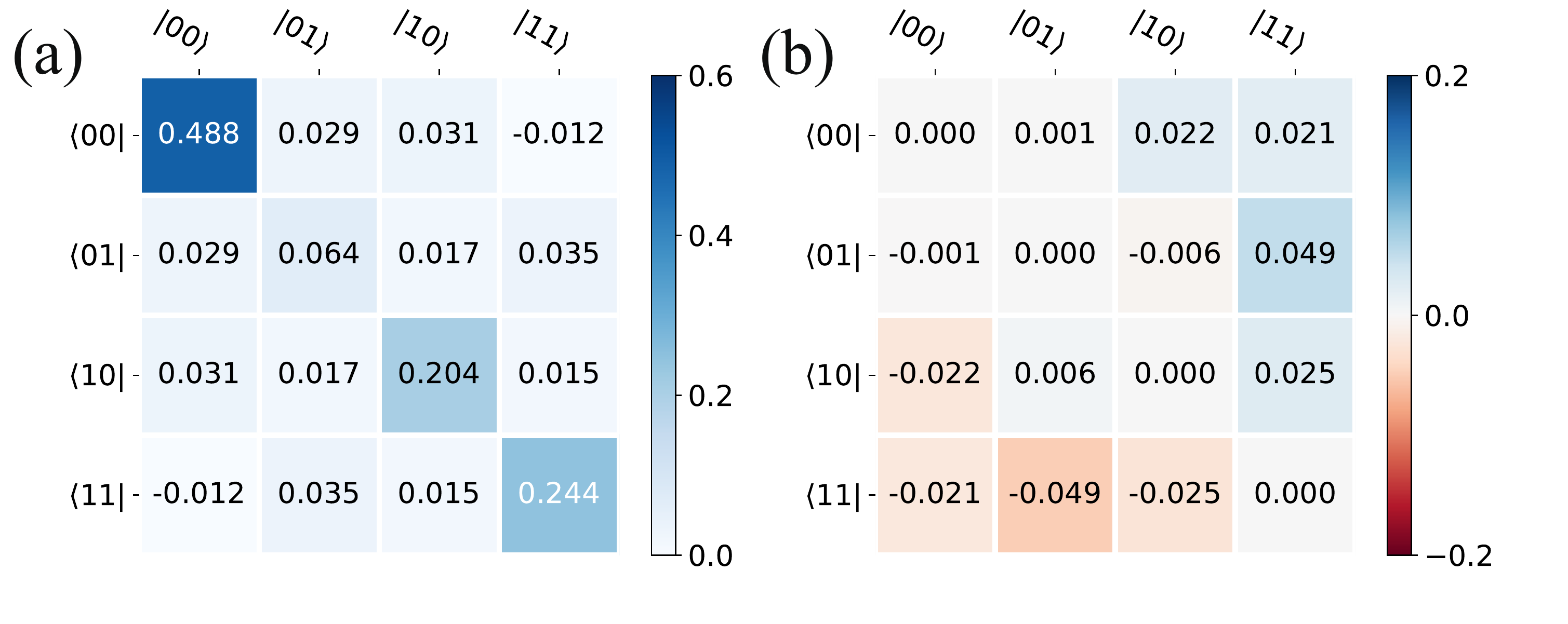}
\caption{(a) real and (b) imaginary part of density matrix of 2-qubit subsystem ${\psi^{(6)}}_{1,5}$  obtained by QOT}
\end{figure}
\begin{figure}[H]
\centering
\includegraphics[width=\linewidth]{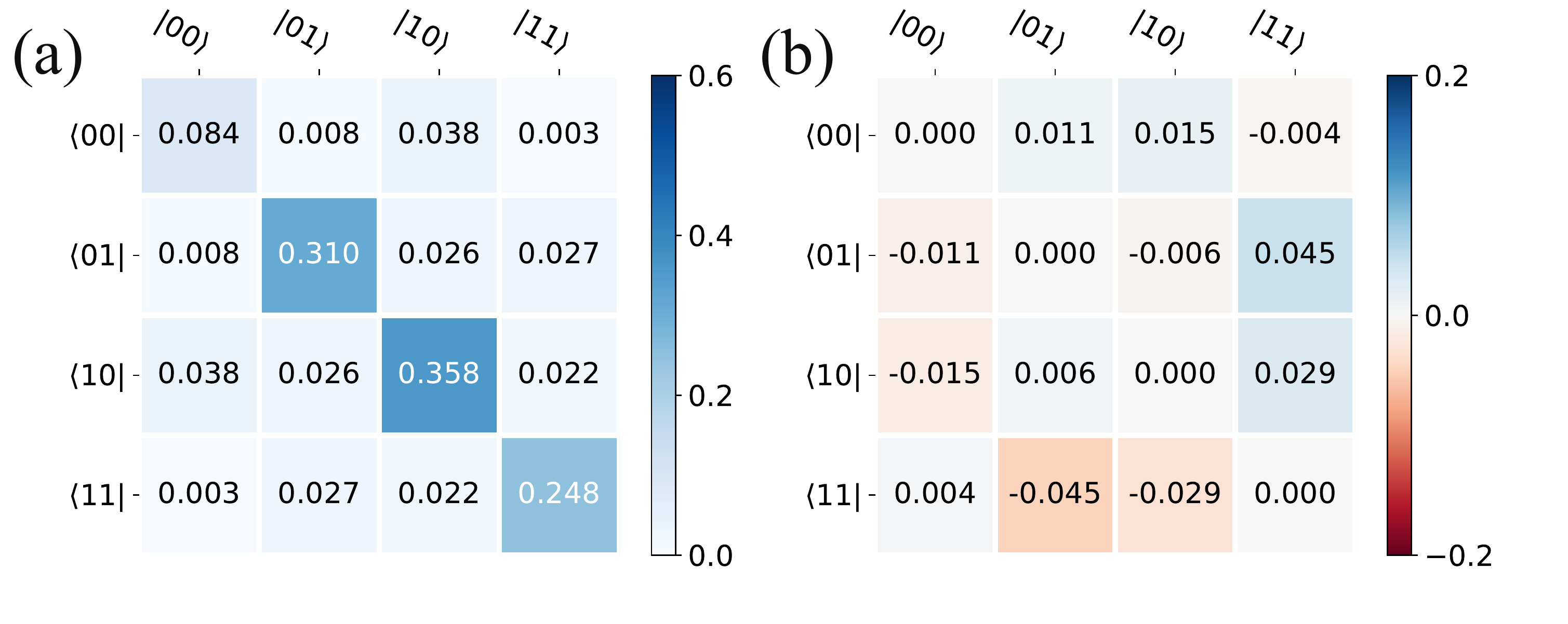}
\caption{(a) real and (b) imaginary part of density matrix of 2-qubit subsystem ${\psi^{(6)}}_{2,3}$  obtained by QOT}
\end{figure}
\begin{figure}[H]
\centering
\includegraphics[width=\linewidth]{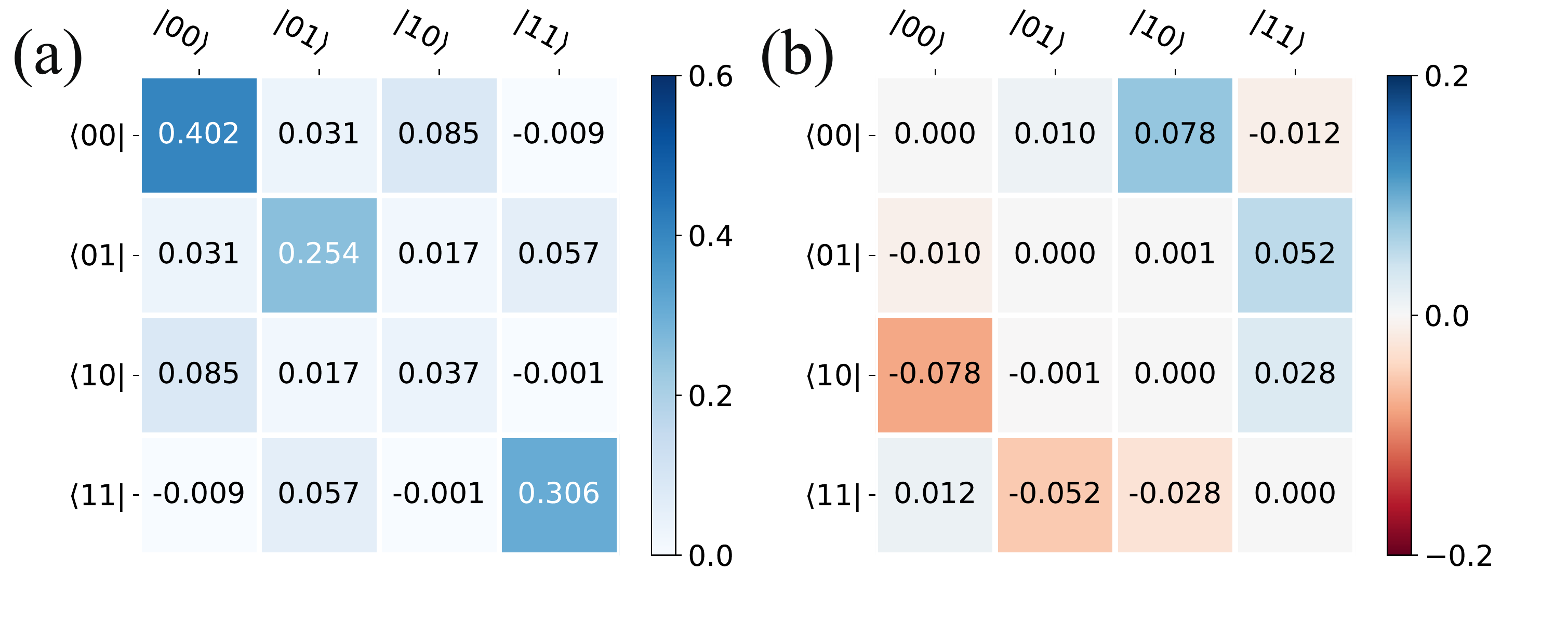}
\caption{(a) real and (b) imaginary part of density matrix of 2-qubit subsystem ${\psi^{(6)}}_{2,4}$  obtained by QOT}
\end{figure}
\begin{figure}[H]
\centering
\includegraphics[width=\linewidth]{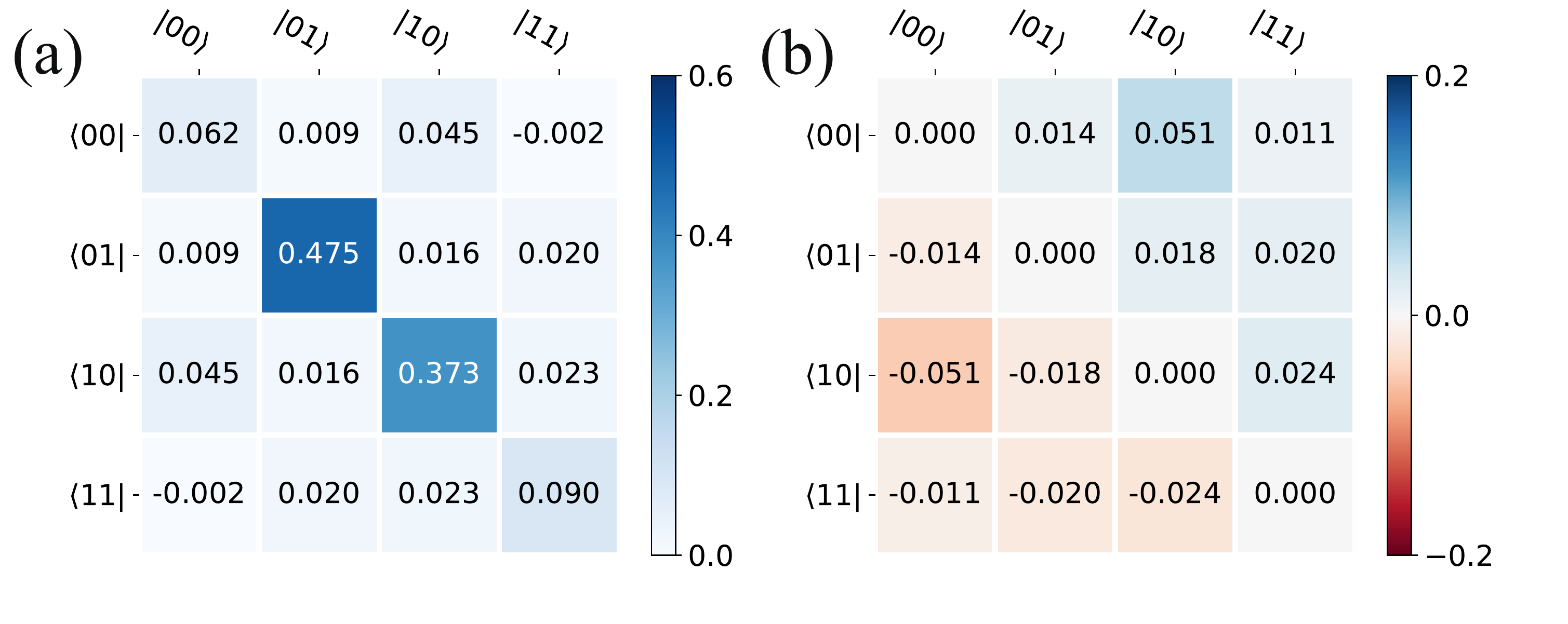}
\caption{(a) real and (b) imaginary part of density matrix of 2-qubit subsystem ${\psi^{(6)}}_{2,5}$  obtained by QOT}
\end{figure}
\begin{figure}[H]
\centering
\includegraphics[width=\linewidth]{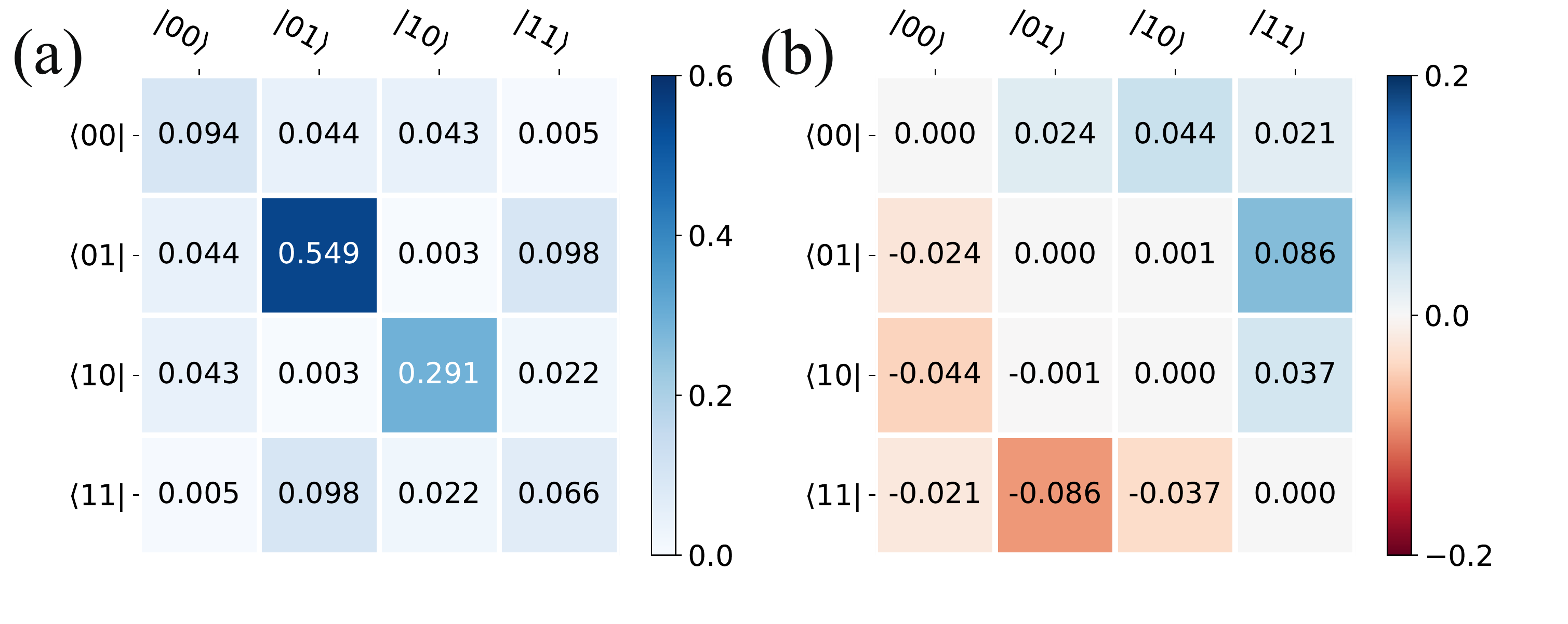}
\caption{(a) real and (b) imaginary part of density matrix of 2-qubit subsystem ${\psi^{(6)}}_{3,4}$  obtained by QOT}
\end{figure}
\begin{figure}[H]
\centering
\includegraphics[width=\linewidth]{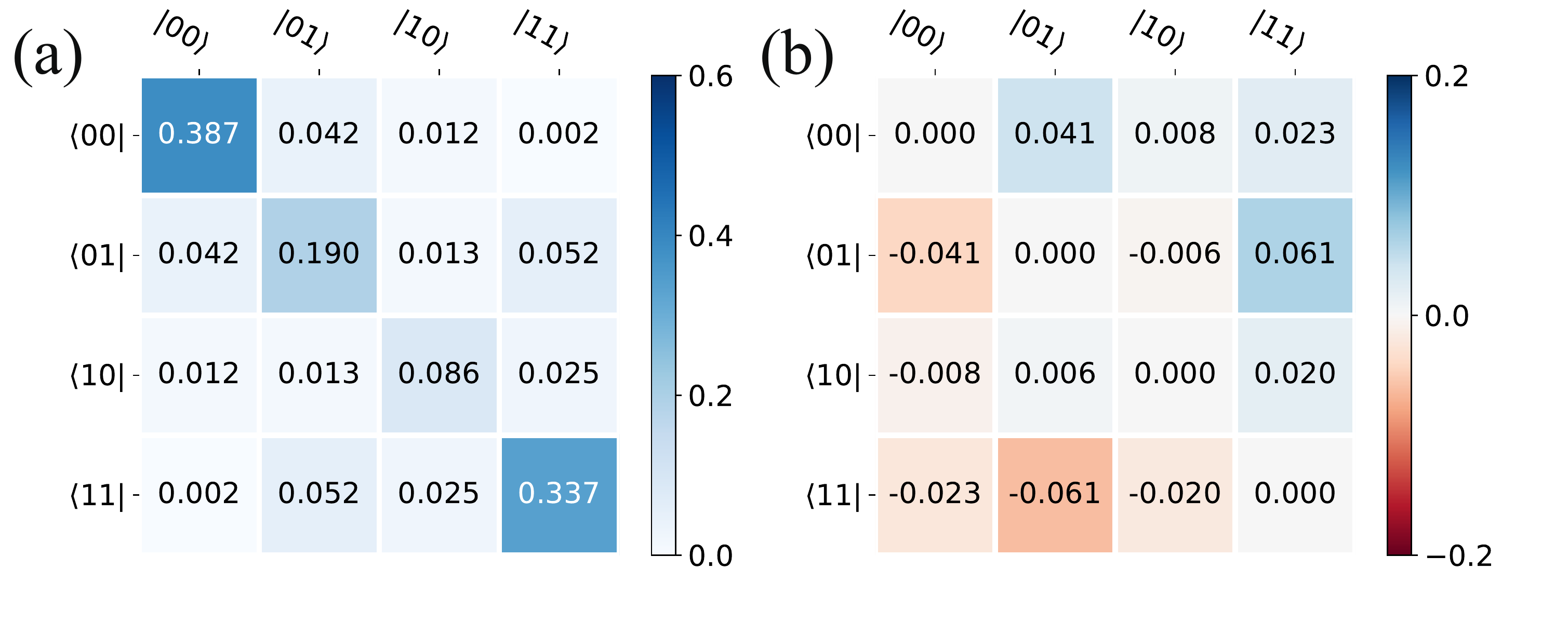}
\caption{(a) real and (b) imaginary part of density matrix of 2-qubit subsystem ${\psi^{(6)}}_{3,5}$  obtained by QOT}
\end{figure}
\begin{figure}[H]
\centering
\includegraphics[width=\linewidth]{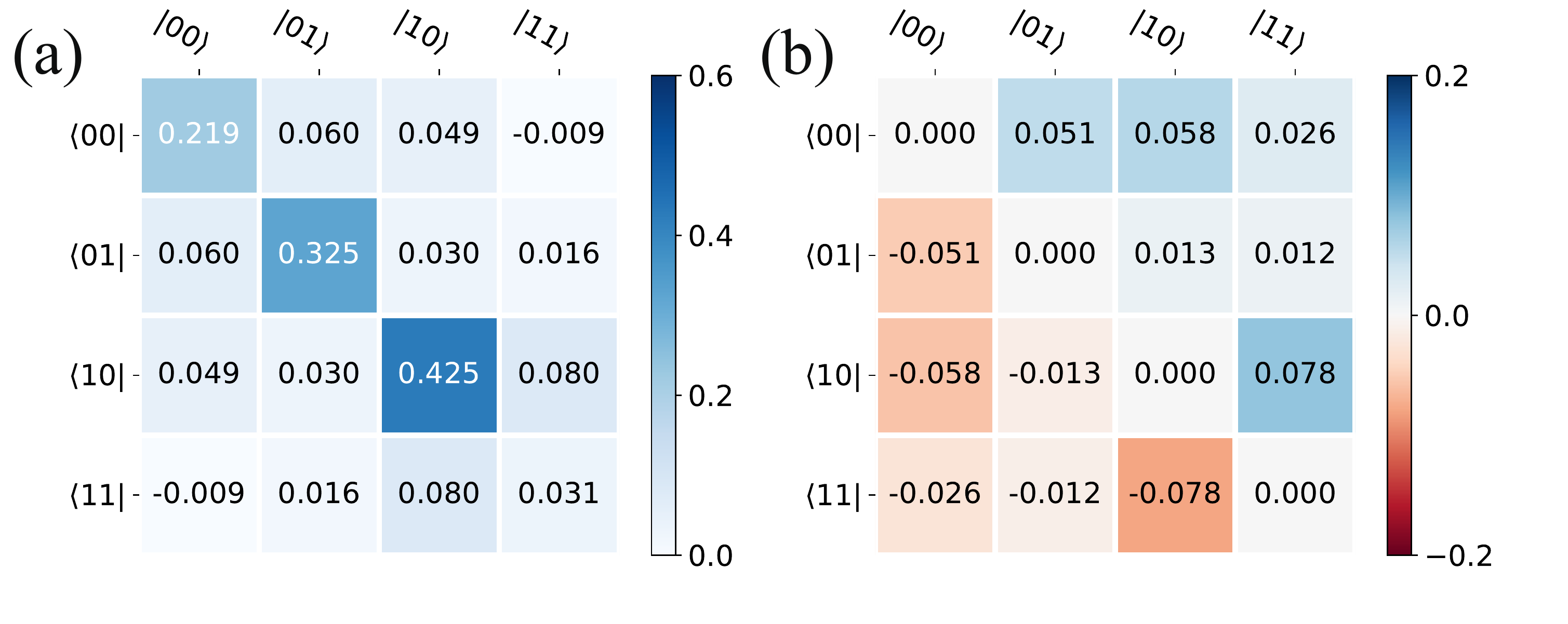}
\caption{(a) real and (b) imaginary part of density matrix of 2-qubit subsystem ${\psi^{(6)}}_{4,5}$  obtained by QOT}
\end{figure}

\newpage

%



